\documentclass{article}

\usepackage[tbtags]{amsmath}
\usepackage{amsfonts,amssymb,mathrsfs,amscd,amsmath,comment}
\usepackage{im2e-tr}
\usepackage{hypbmsec}  
\usepackage[active]{srcltx}
\draft
\theoremstyle{plain}
\newtheorem{theorem}{Theorem}
\newtheorem{lemma}{Lemma}
\newtheorem{corollary}{Corollary}
\newtheorem{proposition}{Proposition}
\newtheorem{prop}{Proposition}

\makeatletter
\renewcommand{\@Alph}[1]{%
  \ifcase#1\or A-1\or A-2\or A-3\else\@ctrerr\fi}
\makeatother

\newtheorem{corol}{Corollary}

\makeatletter
\renewcommand{\@Alph}[1]{%
  \ifcase#1\or A-1\or A-2\or A-3\else\@ctrerr\fi}
\makeatother

\newtheorem{lemm}{Lemma}

\makeatletter
\renewcommand{\@Alph}[1]{%
  \ifcase#1\or A-1\or A-2\or A-3\else\@ctrerr\fi}
\makeatother

\theoremstyle{definition}

\newtheorem{example}{Example}

\theoremstyle{remark}
\newtheorem{remark}{Remark}
\newtheorem{proof}{Proof}

\let\le=\leqslant
\let\ge=\geqslant

\begin{document}
\doi{}
\issyear{2010}
\issvol{74}
\issnum{4}
\isspages{}
\rusisspages{189--224}

\translator{the author}


\author{M.~E.~Shirokov}
\address{Steklov Mathematical Institute, Moscow,
Russia} \email{msh@mi.ras.ru}


\date{16.06.2008 \\ 21.04.2009}

\title{On properties of the space of quantum states and their application to construction of entanglement monotones}

\maketitle

\markright{On properties of the space of quantum states}

\begin{fulltext}

\thanks{This work is partially supported by the program
"Mathematical control theory" of Russian Academy of Sciences, by the
analytical departmental target program "Development of scientific
potential of the higher school 2009-2010" (project 2.1.1/500), by
the federal target program "Scientific and pedagogical staff of
innovative Russia" (program 1.2.1, contract P 938), by RFBR grants
09-01-00424-a and 10-01-00139-a.}

\begin{abstract}
We consider two properties of the set of quantum states as a convex topological space and some their implications concerning the notions of a convex hull and of a convex roof of a function defined on a subset of quantum states.

By using these results we analyze two infinite-dimensional versions (discrete and continuous) of the convex roof construction of entanglement monotones, which is widely used in finite dimensions. It is shown that the discrete version may be 'false' in the sense that the resulting functions may not possess the main property of entanglement monotones while the continuous version can be considered as a 'true' generalized convex roof construction. We give several examples of entanglement monotones produced by this construction. In particular, we consider an infinite-dimensional generalization of the notion of Entanglement of Formation and study its properties.
\end{abstract}


\section*{Introduction}
\label{s0}

In study of finite dimensional quantum systems and channels such
notions of the convex analysis as the convex hull and the convex
closure (called also the convex envelope) of a function defined on
the set of quantum states as well as the convex roof of a function defined on
the set of pure quantum states (introduced in \cite{1} as a special
convex extension of this function to the set of all quantum states)
are widely used. The last notion plays the basic role in
construction of entanglement monotones -- functions on the set of
states of a composite quantum system characterizing entanglement of
these states~\cite{2},~\cite{3}.

The main problem of using these functional constructions in the
infinite dimensional case consists in necessity to apply them to functions with singular properties
such as discontinuity and unboundedness (including possibility of
the infinite values). For instance, the von Neumann entropy -- one
of the main characteristics of quantum states -- is  a continuous
and bounded function in finite dimensions, but it takes the value
$+\infty$ on a dense subset of the set of states of infinite
dimensional quantum system. The other problems are noncompactness of
the set of quantum states and nonexistence of inner points of this
set (considered as a subset of the Banach space of trace class
operators). All these features lead to very "unnatural" behavior of
the above functional constructions and to breaking  validity of
several "elementary" results (for example, the well known Jensen's
inequality may not hold for a measurable convex function). So, a
special analysis is required to overcome these problems. The main
tools of this analysis are the following two properties of the
set of quantum states  as a convex topological space:
\begin{itemize}
\item[1)]
the weak compactness of the set of measures, whose barycenters form a
compact set,
\item[2)] the openness of the barycenter map (in the weak topology),
\end{itemize}
proved in \cite{4} and \cite{5} respectively and described in
detail in~\S\,\ref{s1}. These properties reflect the special relations
between the topology and the convex structure of the set of quantum
states.

In \S\,\ref{s2} the infinite dimensional versions of the notions of
the convex hull of a function defined on the set
$\mathfrak{S}(\mathcal{H})$ and of the convex roof of a function
defined on the set $\operatorname{extr}\mathfrak{S}(\mathcal{H})$
are considered. Their continuity properties are explored. Continuity
of the operation of convex closure with respect to monotone
pointwise convergence on the class of lower semicontinuous lower
bounded functions on $\mathfrak{S}(\mathcal{H})$ is proved.

In \S\,\ref{s3} sufficient conditions for continuity and for
coincidence of restrictions of different convex hulls of a given
function to the set of states with bounded mean generalized energy
(nonnegative lower semicontinuous affine function) are obtained. This result
implies several useful properties of the output Renyi entropy (in
particular, of the output von Neumann entropy) of a quantum channel.

In \S\,\ref{s4} applications of the obtained results to the theory
of entanglement in composite quantum system are considered
\cite{6}. The two infinite dimensional versions (discrete and continuous) of
the convex roof construction
of entanglement monotones widely used in finite dimensions are considered. It is
shown that the discrete version may be "false" in the sense that the
functions constructed by using this method may not possess the main
property of entanglement monotones (even if the generating
function is bounded and lower semicontinuous), while the continuous
version produces "true" entanglement monotones under weak
requirements on the generating functions. So, the last method is
considered as a generalized convex roof construction. It can be
applied to obtain infinite dimensional generalization of
the Entanglement of Formation (EoF) -- one the basic entanglement
measures in finite dimensional composite quantum systems~\cite{7}.
Comparison of this approach to generalization of EoF with the
approach proposed in~\cite{8} is considered in~\S\,\ref{s5}.

\section{Preliminaries}
\label{s1}

Let $\mathcal{H}$ be a separable Hilbert space,
$\mathfrak{B}(\mathcal{H})$ ~--  the algebra of all linear bounded
operators in ~$\mathcal{H}$, $\mathfrak{B}_{h}(\mathcal{H})$~-- the Banach space of
bounded Hermitian operators in ~$\mathcal{H}$ containing the cone
$\mathfrak{B}_{+}(\mathcal{H})$ of positive operators,
$\mathfrak{T}( \mathcal{H})$ and $\mathfrak{T}_{h}(\mathcal{H})$~--
the separable Banach spaces of all trace class operators in ~$\mathcal{H}$ and of all
trace class Hermitian operators with the trace norm
$\|\cdot\|_{1}=\operatorname{Tr}{|\cdot|}$ (cf.~\cite{9}).

The closed subsets
$$
\mathfrak{T}_{1}(\mathcal{H})=
\bigl\{A\in\mathfrak{T}(\mathcal{H})\mid A\ge0,\operatorname{Tr}A\le1\bigr\},
\qquad
\mathfrak{S}(\mathcal{H})=\bigl\{A\in\mathfrak{T}_{1}(\mathcal{H})\mid \operatorname{Tr}A=1\bigr\}
$$
of  $\mathfrak{T}(\mathcal{H})$ are  complete separable
metric spaces with the metric defined by the trace norm. An operator
$\rho$ in $\mathfrak{S}(\mathcal{H})$ determines the linear
functional $A\mapsto\operatorname{Tr}A\rho$ on the algebra
$\mathfrak{B}(\mathcal{H})$ called \textit{state} in the theory of
operator algebras. So, in what follows, we will use the term
\textit{state} for operators in $\mathfrak{S}(\mathcal{H})$. The
\textit{rank} of a positive operator (state) is the dimension of the
orthogonal complement of its kernel.

We will denote by $\operatorname{co}\mathcal{A}$ (correspondingly,
$\overline{\operatorname{co}}\mathcal{A}$) the convex hull
(correspondingly, closure) of a set $\mathcal{A}$\enskip
\cite{10}. We will denote by
$\operatorname{extr}\mathcal{A}$ the set of all extreme points of a
convex set $\mathcal{A}$.

We will denote by $\mathcal{P}(\mathcal{A})$ the set of all Borel
probability measures on a complete separable metric space
$\mathcal{A}$ endowed with the topology of weak convergence. This
set can be considered as a complete separable metric space as well
\cite{11}, Ch.\,II, \S\,6. The subset of $\mathcal{P}(\mathcal{A})$
consisting of measures with finite support will be denoted
$\mathcal{P}^{\mathrm{f}}(\mathcal{A})$. In what follows we will
also use the abbreviations
$\mathcal{P}=\mathcal{P}(\mathfrak{S}(\mathcal{H}))$,
$\widehat{\mathcal{P}}=\mathcal{P}(\operatorname{extr}
\mathfrak{S}(\mathcal{H}))$.

The \textit{barycenter} of the measure $\mu\in\mathcal{P}$~is the
state defined by the
Bochner integral
$$
\bar{\rho}(\mu)=\int_{\mathfrak{S}(\mathcal{H})}\sigma \mu(d\sigma).
$$

For an arbitrary subset $\mathcal{A}\subset\mathfrak{S}(\mathcal{H})$
denote by $\mathcal{P}_{\mathcal{A}}$ (correspondingly, by
$\widehat{\mathcal{P}}_{\mathcal{A}}$) the subset of
$\mathcal{P}$ (correspondingly, of  $\widehat{\mathcal{P}}$),
consisting of all measures with the barycenter in~$\mathcal{A}$.

A finite or countable collection of states $\{\rho_{i}\}$ with
corresponding probability distribution $\{\pi_{i}\}$ is
conventionally called \textit{ensemble} and is denoted $\{\pi
_{i},\rho _{i}\}$.  In this paper we will consider ensemble of
states as a particular case of probability measure on the set of
quantum states.

The \textit{von Neumann entropy} of a state $\rho$ and the
\textit{relative entropy} of states $\rho$ and  $\sigma$ are defined
respectively by the expressions
$$
H(\rho)=-\sum_{i}\langle i|\,\rho\log \rho\,|i\rangle,
\qquad
H(\rho\,\|\sigma)=\sum_{i}\langle i|\,(\rho\log \rho-\rho\log \sigma)\,|i\rangle,
$$
where $\{|i\rangle\}$~is a basic of eigenvectors of $\rho$, and it
is assumed that $H(\rho\,\|\sigma)=+\infty$ if the support of $\rho$
(the orthogonal complement of the kernel of the operator $\rho$) is
not contained within the support of the state $\sigma$. The entropy
and the relative entropy are lower semicontinuous functions of their
arguments taking values in~$[0,+\infty]$. The first of them is
concave while the second one is jointly convex \cite{12}.

An arbitrary positive unbounded operator $H$ in a
space $\mathcal{H}$ with discrete spectrum of finite multiplicity
will be called $\mathfrak{H}$-\textit{operator}.

The set of quantum states $\mathfrak{S}(\mathcal{H})$ has the
following two properties:

A) for an arbitrary compact subset
$\mathcal{A}\subset\mathfrak{S}(\mathcal{H})$ the set
$\mathcal{P}_{\mathcal{A}}(\mathfrak{S}(\mathcal{H}))$ is compact
(see \cite{4});

B) the barycenter map
$\mathcal{P}(\mathfrak{S}(\mathcal{H}))\ni\mu\mapsto\bar{\rho}(\mu)\in\mathfrak{S}(\mathcal{H})$
is an open surjection (see \cite{5},~\cite{13}).

Property A) provides generalization to the case of
$\mathfrak{S}(\mathcal{H})$ of some well known results concerning
compact convex sets (see~\cite{14}, Lemma 1, or the below
Propositions \ref{p1} and \ref{p6}) and hence it may be considered
as a kind of "weak" compactness. In fact, this property is not purely
topological (in contrast to compactness), but it reflects the special
relation between the topology and the convex structure of the set
$\mathfrak{S}(\mathcal{H})$. Following \cite{13}, \cite{15}, we will call it the \textit{$\mu$-compactness
property}.

Note that the $\mu$-compactness of the positive part of the unit
ball is a specific feature of the Banach space of trace class
operators (the Shatten class of order $p=1$) within the family of
Shatten classes of order $p\ge1$.

Moreover, it can be shown that the set
$\mathfrak{T}_{1}(\mathcal{H})$ loses the $\mu$-compactness property
being endowed with the $\|\cdot\|_{p}$-norm topology with $p>1$ and
that in the Shatten class of order $p=2$ (the Hilbert space of
Hilbert-Schmidt operators) there exists no $\mu$-compact set which
is not compact. These and other results concerning the
$\mu$--compactness property as well as examples of $\mu$-compact
sets are considered in~\cite{15}.

Property B) reflects an another relation between the topology and the
convex structure of the set $\mathfrak{S}(\mathcal{H})$. The
characterization of the analog of this property for arbitrary
$\mu$-compact convex set is obtained in ~\cite{13}, Theorem\,1.\footnote{This theorem is a partial noncompact generalization
of the results in \cite{16}, concerning compact convex sets. The
complete generalization of these results to the class of
$\mu$\nobreakdash-\hspace{0pt}compact convex sets is obtained
in~\cite{15}.} By this theorem property B) is equivalent to
continuity of the convex hull of any continuous bounded function on
the set $\mathfrak{S}(\mathcal{H})$ and to openness of the map
$$
\mathfrak{S}(\mathcal{H})\times\mathfrak{S}(\mathcal{H})\times[0,1]\ni (\rho,\sigma,\lambda)
\mapsto\lambda  \rho + (1-\lambda)\sigma \in \mathfrak{S}(\mathcal{H}).
$$
The analog of the last property for any convex set  seems to
be the simplest for verification and (its equivalent but formally
stronger form) is called the \textit{stability} property (see
~\cite{17}, \cite{18} and references therein).

\section{The convex hulls and the convex roofs}
\label{s2}

In this section we consider several notions and constructions for
functions defined on the set $\mathfrak{S}(\mathcal{H})$. Note that
the all definitions are universal, they can be formulated in
terms of functions defined on a convex closed bounded subset
$\mathcal{A}$ of a locally convex space (instead of
$\mathfrak{S}(\mathcal{H})$). So, the main results obtained in this
section can be proved in this extended context under the particular
conditions  imposed on $\mathcal{A}$ (which are valid for
$\mathfrak{S}(\mathcal{H})$). Possibilities of such generalizations
are discussed in the Appendix.

\subsection{Several notions of convexity of a function}
\label{s2.1} In what follows we will consider functions on the set
$\mathfrak{S}(\mathcal{H})$ taking values in~$[-\infty,+\infty]$,
which are \textit{semibounded} (lower or upper bounded) on this set.

We will use the following two strengthened versions of the well
known notion of a convex function.

A semibounded function $f$ on the set $\mathfrak{S}(\mathcal{H})$ is
called \textit{$\sigma$-convex} if
$$
f\biggl(\sum_{i}\pi_{i}\rho_{i}\biggr)\le\sum_{i}\pi_{i}f(\rho_{i})
$$
for any \textit{countable} ensemble $\{\pi_{i},\rho_{i}\}$ of states
in $\mathfrak{S}(\mathcal{H})$.

A semibounded universally measurable\footnote{This means that the
function $f$ is measurable with respect to any measure in
$\mathcal{P}(\mathfrak{S}(\mathcal{H}))$\enskip
 \cite{19}.} function $f$ on the set
$\mathfrak{S}(\mathcal{H})$ is called \textit{$\mu$-convex} if
$$
f\biggl(\int_{\mathfrak{S}(\mathcal{H})}\rho\mu(d\rho)\biggr)
\le\int_{\mathfrak{S}(\mathcal{H})}f(\rho)\mu(d\rho)
$$
for any measure $\mu$ in $\mathcal{P}(\mathfrak{S}(\mathcal{H}))$.

The simplest example of a convex Borel function on the set
$\mathfrak{S}(\mathcal{H})$, which is not $\sigma$-convex and
$\mu$-convex, is the function taking the value $0$ on the convex set
of finite rank states and the value $+\infty$ on set of infinite
rank states. Difference between the above convexity properties can
be also illustrated by functions in the below examples \ref{e1},
\ref{e2} (the first of them is convex but not $\sigma$-convex while
the second one is $\sigma$-convex but not $\mu$-convex).

Convexity implies $\sigma$-convexity for all upper bounded
functions on $\mathfrak{S}(\mathcal{H})$ (Proposition ~\ref{pA-1} in
the Appendix).

By the integral Jensen's inequality (Proposition~\ref{pA-2} in the
Appendix) all these convexity properties are equivalent for the
classes of lower semicontinuous functions and of upper bounded upper
semicontinuous functions on the set $\mathfrak{S}(\mathcal{H})$.

\subsection{The convex hulls and the convex closure}
\label{s2.2} The \textit{convex hull} $\operatorname{co}f$ of a
semibounded function $f$ on the set $\mathfrak{S}(\mathcal{H})$ is
defined as the greatest convex function majorized by $f$\enskip
\cite{20}, which means that
\begin{equation}
\label{eq1}
\operatorname{co}f(\rho)=\inf_{\{\pi_{i},\rho_{i}\}\in\mathcal{P}^{\mathrm{f}}_{\{\rho\}}}
\sum_{i}\pi_{i}f(\rho_{i}),
\qquad
\rho\in\mathfrak{S}(\mathcal{H})
\end{equation}
(the infimum is over all finite ensembles
$\{\pi_{i},\rho_{i}\}$ of states with the average state $\rho$).

The \textit{$\sigma$-convex hull}
$\sigma\textup{-}\!\operatorname{co}f$ of a semibounded function $f$
on the set $\mathfrak{S}(\mathcal{H})$ is defined as follows
\begin{equation}
\label{eq2}
\sigma\textup{-}\!\operatorname{co}f(\rho)=\inf_{\{\pi_{i},\rho_{i}\}\in\mathcal{P}_{\{\rho\}}}
\sum_{i}\pi_{i}f(\rho_{i}),
\qquad
\rho\in\mathfrak{S}(\mathcal{H})
\end{equation}
(the infimum is over all countable ensembles
$\{\pi_{i},\rho_{i}\}$ of states with the average state $\rho$). The
function $\sigma\textup{-}\!\operatorname{co}f$ is $\sigma$-convex,
since for any countable ensemble $\{\lambda_{i},\sigma_{i}\}$ with
the average state $\sigma$ and any family
$\{\{\pi_{ij},\rho_{ij}\}_{j}\}_{i}$ of countable ensembles such
that $\sigma_{i}=\sum_{j}\pi_{ij}\rho_{ij}$ for all $i$ the
countable ensemble $\{\lambda_{i}\pi_{ij},\rho_{ij}\}_{ij}$ has the
average state $\sigma$. Thus $\sigma\textup{-}\!\operatorname{co}f$
is the greatest $\sigma$-convex function majorized by $f$.

The \textit{$\mu$-convex hull} $\mu\textup{-}\!\operatorname{co}f$
of a Borel semibounded function $f$ on the set
$\mathfrak{S}(\mathcal{H})$ is defined as follows
\begin{equation}
\label{eq3}
\mu\textup{-}\!\operatorname{co}f(\rho)=\inf_{\mu\in\mathcal{P}_{\{\rho\}}}
\int_{\mathfrak{S}(\mathcal{H})}f(\sigma)\mu(d\sigma),
\qquad
\rho\in\mathfrak{S}(\mathcal{H})
\end{equation}
(the infimum is over all probability measures $\mu$ with the
barycenter $\rho$). If the function
$\mu\textup{-}\!\operatorname{co}f$ is universally
measurable\footnote{By using the results in \cite{19} universal
measurability of the function $\mu\textup{-}\!\operatorname{co}f$
can be proved for any bounded Borel function $f$.} and $\mu$-convex
then it is the greatest $\mu$-convex function majorized by $f$. By
Propositions \ref{p1} and \ref{p2}  below (used with evident
convexity of the function $\mu\textup{-}\!\operatorname{co}f$ and
Proposition~\ref{pA-2} in the Appendix) this holds if the function
$f$ is either lower bounded and lower semicontinuous or upper bounded and upper
semicontinuous.

The \textit{convex closure} $\overline{\operatorname{co}}f$ of a
lower bounded function $f$ on the set $\mathfrak{S}(\mathcal{H})$ is
defined as the greatest convex lower semicontinuous (closed)
function majorized by $f$\enskip \cite{20}. By Fenchel's
theorem (see~\cite{10}, \cite{20}, \cite{21}) the function
$\overline{\operatorname{co}}f$ coincides with the double Fenchel
transformation of the function $f$, which means that\footnote{To
obtain the below expression from the Fenchel theorem it is necessary
to consider the extension $\hat{f}$ of the function $f$ to the real
Banach space $\mathfrak{T}_{h}(\mathcal{H})$ by setting
$\hat{f}=+\infty$ on
$\mathfrak{T}_{h}(\mathcal{H})\setminus\mathfrak{S}(\mathcal{H})$
and to use coincidence of the space $\mathfrak{B}_{h}(\mathcal{H})$
with the dual space to $\mathfrak{T}_{h}(\mathcal{H})$.}
\begin{equation}
\label{eq4}
\overline{\operatorname{co}}f(\rho)=f^{**}(\rho)
=\sup_{A \in\mathfrak{B}_{+}(\mathcal{H})}[\operatorname{Tr}A\rho-f^{*}(A)],
\qquad
\rho\in\mathfrak{S}(\mathcal{H}),
\end{equation}
where
$$
f^{*}(A)=\sup_{\rho \in \mathfrak{S}(\mathcal{H})}[\operatorname{Tr}A\rho -f(\rho)],
\qquad
A\in\mathfrak{B}_{+}(\mathcal{H}).
$$

It follows from the definitions and Proposition~\ref{pA-2} in the
Appendix that
$$
\overline{\operatorname{co}}f(\rho)\le\mu\textup{-}\!\operatorname{co}f(\rho)
\le\sigma\textup{-}\!\operatorname{co}f(\rho)\le\operatorname{co}f(\rho),
\qquad
\rho\in\mathfrak{S}(\mathcal{H}),
$$
for any  Borel lower bounded function $f$ on the set
$\mathfrak{S}(\mathcal{H})$. It is possible to prove (see Corollary
 \ref{c1} below) that the equalities hold in the above
inequalities for any continuous bounded function $f$ on the set
$\mathfrak{S}(\mathcal{H})$. The following examples show that the
last assertion is not true in general.
\begin{example}
\label{e1} Let $H$ be the von Neumann entropy (see\ \S\,\ref{s1})
and $\rho_{0}$ be a state such that $H(\rho_{0})=+\infty$. Since the
set of quantum states with finite entropy is convex,
$\operatorname{co}H(\rho_{0})=+\infty$ while the spectral theorem
implies $\sigma\textup{-}\!\operatorname{co}H(\rho_{0})=0$.
\end{example}

\begin{example}
\label{e2} Let $f$ be the indicator function of the complement of
the closed set $\mathcal{A}_{s}$ of pure product states in
$\mathfrak{S}(\mathcal{H}\otimes\mathcal{H})$ and $\omega_{0}$ be
the separable state in $\overline{\operatorname{co}}\mathcal{A}_{s}$
constructed in~\cite{14} such that any measure in
$\mathcal{P}_{\{\omega_{0}\}}(\mathfrak{S}(\mathcal{H}\otimes\mathcal{H}))$
have no atoms in $\mathcal{A}_{s}$. It is easy to show that
$\sigma\textup{-}\!\operatorname{co}f(\omega_{0})=1$. By Lemma\,1 in \cite{14} there exists a measure $\mu_{0}$ in
$\mathcal{P}_{\{\omega_{0}\}}(\mathfrak{S}(\mathcal{H}\otimes\mathcal{H}))$
supported by the set $\mathcal{A}_{s}$. Hence
$\mu\textup{-}\!\operatorname{co}f(\omega_{0})=0$. Note that
$\sigma\textup{-}\!\operatorname{co}f$~is a $\mu_{0}$-integrable
$\sigma$-convex bounded function on the set
$\mathfrak{S}(\mathcal{H}\otimes\mathcal{H})$, for which Jensen's
inequality does not hold:
$$
1=\sigma\textup{-}\!\operatorname{co}f(\omega_{0})
>\int_{\mathfrak{S}(\mathcal{H}\otimes\mathcal{H})}
\sigma\textup{-}\!\operatorname{co}f(\omega)\mu_{0}(d\omega)=0
$$
(since the functions $\sigma\textup{-}\!\operatorname{co}f$ and $f$
coincide on the support of the measure $\mu_{0}$).
\end{example}

\begin{example}
\label{e3} Let $f$ be the indicator function of a set consisting of
one pure state. Then $\mu\textup{-}\!\operatorname{co}f=f$ while
$\overline{\operatorname{co}}f\equiv 0$.
\end{example}

Since the set $\mathfrak{S}(\mathcal{H})$ is $\mu$-compact,
Proposition\,3 in \cite{13} implies the following assertion.

\begin{proposition}
\label{p1} Let $f$ be a lower semicontinuous lower bounded function
$f$ on the set $\,\mathfrak{S}(\mathcal{H})$. Then the $\mu$-convex
hull of this function is lower semicontinuous, which means that
\begin{equation}
\label{eq5}
\overline{\operatorname{co}}f(\rho)=\mu\textup{-}\!\operatorname{co}f(\rho)
=\inf_{\mu\in\mathcal{P}_{\{\rho\}}}\int_{\mathfrak{S}(\mathcal{H})}f(\sigma)\mu(d\sigma),
\qquad
\rho\in\mathfrak{S}(\mathcal{H}).
\end{equation}
The infimum in~\eqref{eq5} is achieved at some measure in
$\mathcal{P}_{\{\rho\}}$.
\end{proposition}

The $\mu$-compactness of the set $\mathfrak{S}(\mathcal{H})$ is an
essential condition of validity of representation \eqref{eq5} for
the convex closure \cite{15}, Proposition\,7. Representation
\eqref{eq5} implies, in particular, that the convex closure of an
arbitrary lower semicontinuous lower bounded function on the set
$\mathfrak{S}(\mathcal{H})$ coincides with this function on the set
$\operatorname{extr}\mathfrak{S}(\mathcal{H})$ of pure states.

Note also that the condition of lower boundedness in Proposition
\ref{p1} is essential, since Lemma \ref{l2} below shows that if a convex lower
semicontinuous function is not lower bounded on the set
$\mathfrak{S}(\mathcal{H})$~ then it is equal to~$-\infty$ everywhere.

Stability of the set $\mathfrak{S}(\mathcal{H})$ implies the
following result.

\begin{proposition}
\label{p2} Let $f$ be an upper semicontinuous function on the set
$\mathfrak{S}(\mathcal{H})$. Then the convex hull $\,\mathrm{co}f$
of this function is upper semicontinuous. If, in addition, the
function $f$ is upper bounded then the convex hull, the
$\sigma$-convex hull and the $\mu$-convex hull of this function
coincide:
$\operatorname{co}f=\sigma\textup{-}\!\operatorname{co}f=\nobreak\mu\textup{-}\!\operatorname{co}f$.
\end{proposition}

\begin{proof}
Upper semicontinuity of the function $\operatorname{co}f$ can be
proved by using the more general assertion of Lemma \ref{l4} below, since
for an arbitrary sequence $\{\rho_{n}\}$ of states in
$\mathfrak{S}(\mathcal{H})$, converging to a state $\rho_{0}$,
Lemma\,3 in \cite{4} implies existence of such $\mathfrak{H}$-operator
$H$ in the space $\mathcal{H}$ that
$\sup_{n\ge0}\operatorname{Tr}H\rho_{n}<+\infty$.

Coincidence of the functions  $\operatorname{co}f$ and
$\mu\textup{-}\!\operatorname{co}f$ under the condition of upper
boundedness of the function $f$ is easily proved by using upper
semicontinuiuty of the functional $\mu\mapsto\nobreak
\int_{\mathfrak{S}(\mathcal{H})}f(\rho)\mu(d\rho)$ on the set
$\mathcal{P}(\mathfrak{S}(\mathcal{H}))$ and density of measures
with finite support in the set of all measures with given
barycenter~\cite{4}, Lemma\,1.

Example \ref{e3} shows that the condition of Proposition \ref{p2}
does not imply coincidence of the function
$\overline{\operatorname{co}}f$ with the function
$\mu\textup{-}\!\operatorname{co}f=\sigma\textup{-}\!\operatorname{co}f=\operatorname{co}f$.
\end{proof}

Propositions \ref{p1} and \ref{p2} have the following obvious
corollary.

\begin{corollary}
\label{c1} Let $f$ be a continuous lower bounded function on the set
$\,\mathfrak{S}(\mathcal{H})$. Then the convex hull
$\,\operatorname{co}f$ is continuous on any subset of
$\,\mathfrak{S}(\mathcal{H})$, where it coincides with the
$\mu$-convex hull $\,\mu\textup{-}\!\operatorname{co}f$.

If, in addition, the function $f$ is  bounded then its convex hull,
$\sigma$-convex hull, $\mu$-convex hull, convex closure coincide:
$\operatorname{co}f=\sigma\textup{-}\!\operatorname{co}f
=\mu\textup{-}\!\operatorname{co}f=\overline{\operatorname{co}}f$
and this function is continuous.
\end{corollary}

By using Proposition \ref{p1} it is easy to show that a necessary
and sufficient condition of coincidence of the functions
$\operatorname{co}f$ and $\mu\textup{-}\!\operatorname{co}f$ at a
state $\rho_{0}\in\mathfrak{S}(\mathcal{H})$ consists in validity of
the Jensen inequality $\operatorname{co}f(\rho_{0})\leq
\int\operatorname{co}f(\rho)\mu(d\rho)$ for any measure $\mu$ in
$\mathcal{P}_{\{\rho_{0}\}}$ (the convex function
$\operatorname{co}f$ is Borel by Proposition \ref{p2}). The
particular sufficient condition of this coincidence is considered in
Corollary \ref{c6} below.

The second assertion of Corollary \ref{c1} shows that
\begin{equation}
\label{eq6}
\overline{\operatorname{co}}f(\rho)
=\operatorname{co}f(\rho)=\inf_{\{\pi_{i},\rho_{i}\}\in\mathcal{P}^{\mathrm{f}}_{\{\rho\}}}
\sum_{i}\pi_{i}f(\rho_{i}),
\qquad
\rho\in\mathfrak{S}(\mathcal{H}),
\end{equation}
for any continuous bounded function $f$ on the set
$\mathfrak{S}(\mathcal{H})$. This representation for a convex closure
is a noncompact generalization of Corollary I.3.6 in~\cite{22}.

We will use the following approximation result.

\begin{lemma}
\label{l1} Let $f$ be a Borel lower bounded function on the set
$\mathfrak{S}(\mathcal{H})$.  For an arbitrary state $\rho_{0}$ in
$\mathfrak{S}(\mathcal{H})$ there exists a sequence $\{\rho_{n}\}$,
converging to the state $\rho_{0}$, such that
$$
\limsup_{n\rightarrow+\infty}\sigma\textup{-}\!\operatorname{co}f(\rho_{n})
\le\limsup_{n\rightarrow+\infty}\operatorname{co}f(\rho_{n})
\le\mu\textup{-}\!\operatorname{co}f(\rho_{0}).
$$
If, in addition, the function $f$ is lower semicontinuous then
$$
\lim_{n\rightarrow+\infty}\sigma\textup{-}\!\operatorname{co}f(\rho_{n})
=\lim_{n\rightarrow+\infty}\operatorname{co}f(\rho_{n})
=\mu\textup{-}\!\operatorname{co}f(\rho_{0}).
$$
\end{lemma}

\begin{proof}
It is sufficient to consider the case of nonnegative function $f$.
For given natural $n$ let $\mu_{n}$ be a measure in
$\mathcal{P}_{\{\rho_{0}\}}$ such that
$$
\mu\textup{-}\!\operatorname{co}f(\rho_{0})\ge
\int_{\mathfrak{S}(\mathcal{H})}f(\rho)\mu_{n}(d\rho)-\frac1n.
$$
Since the set $\mathfrak{S}(\mathcal{H})$ is separable there exists a
sequence $\{\mathcal{A}_{i}^{n}\}$ of Borel subsets of
$\mathfrak{S}(\mathcal{H})$ with diameter $\leq 1/n$ such that
$\mathfrak{S}(\mathcal{H})=\bigcup_{i}\mathcal{A}_{i}^{n}$ and
$\mathcal{A}_{i}^{n}\cap \mathcal{A}_{j}^{n}=\varnothing$ if $j\neq
i$. Let $m=m(n)$ be such number that $\sum_{i=m+1}^{+\infty }\mu_{n}
(\mathcal{A}_{i}^{n})<1/n$. Without loss of generality we may assume
that $\mu_{n}(\mathcal{A}^{n}_{i})>0$ for $i=\overline{1,m}$. For
each  $i$ the set $\mathcal{A}^{n}_{i}$ contains a state
$\rho^{n}_{i}$ such that $f(\rho^{n}_{i})\le
(\mu_{n}(\mathcal{A}^{n}_{i}))^{-1}\int_{\mathcal{A}^{n}_{i}}f(\rho)\mu_{n}(d\rho)$.

Let $\mathcal{B}_{n}\,{=}\,\bigcup_{i=1}^{m}\mathcal{A}^{n}_{i}$.
Consider the state
$\rho_{n}\,{=}\,(\mu_{n}(\mathcal{B}_{n}))^{-1}\sum_{i=1}^{m}\mu_{n}
(\mathcal{A}^{n}_{i})\rho^{n}_{i}$. We want to show that
\begin{equation}
\label{eq7}
\lim_{n\rightarrow+\infty}\rho_{n}=\rho_{0}.
\end{equation}

For each $i$ the state
$\hat{\rho}^{n}_{i}=(\mu_{n}(\mathcal{A}^{n}_{i}))^{-1}\int_{\mathcal{A}^{n}_{i}}\rho\mu_{n}(d\rho)$
lies in the set $\overline{\operatorname{co}}(\mathcal{A}^{n}_{i})$
with diameter $\leq 1/n$. It follows that
$\|\rho^{n}_{i}-\hat{\rho}^{n}_{i}\|_{1}\le 1/n$ for
$i=\overline{1,m}$. By noting that
$\mu_{n}(\mathcal{B}_{n})=\sum_{i=1}^{m}\mu_{n}(\mathcal{A}^{n}_{i})$,
we have
\begin{align*}
\|\rho_{n}-\rho_{0}\|_{1}
&=\biggl\|(\mu_{n}(\mathcal{B}_{n}))^{-1}\sum_{i=1}^{m}\mu_{n}(\mathcal{A}^{n}_{i})
\rho^{n}_{i}-\sum_{i=1}^{m}\int_{\mathcal{A}^{n}_{i}}\rho\mu_{n}(d\rho)-
\int_{\mathfrak{S}(\mathcal{H})\setminus\mathcal{B}_{n}}\rho\mu_{n}(d\rho)\biggr\|_{1}
\\
&\le \sum_{i=1}^{m}\mu_{n}(\mathcal{A}^{n}_{i})
\|(\mu_{n}(\mathcal{B}_{n}))^{-1}\rho^{n}_{i}-\hat{\rho}^{n}_{i}\|_{1}
+\biggl\|\int_{\mathfrak{S}(\mathcal{H})\setminus\mathcal{B}_{n}}\rho\mu_{n}(d\rho)\biggr\|_{1}
\\
&\le(1-\mu_{n}(\mathcal{B}_{n}))+\sum_{i=1}^{m}\mu_{n}(\mathcal{A}^{n}_{i})
\|\rho^{n}_{i}-\hat{\rho}^{n}_{i}\|_{1}+
\mu_{n}(\mathfrak{S}(\mathcal{H})\setminus\mathcal{B}_{n})<\frac3n,
\end{align*}
which implies \eqref{eq7}.

By the choice of the states $\rho^{n}_{i}$ we have
\begin{align*}
\operatorname{co}f(\rho_{n})
&\le(\mu_{n}(\mathcal{B}_{n}))^{-1}\sum_{i=1}^{m}\mu_{n}(\mathcal{A}^{n}_{i})f(\rho^{n}_{i})
\le(\mu_{n}(\mathcal{B}_{n}))^{-1}\sum_{i=1}^{m}\int_{\mathcal{A}^{n}_{i}}f(\rho)\mu_{n}(d\rho)
\\
&\le(\mu_{n}(\mathcal{B}_{n}))^{-1}\int_{\mathfrak{S}(\mathcal{H})}
f(\rho)\mu_{n}(d\rho)
\le\biggl(1-\frac1n\biggr)^{-1}\biggl(\mu\textup{-}\!\operatorname{co}f(\rho_{0})+\frac 1n\biggr).
\end{align*}
This implies the first assertion of the lemma. By
Proposition~\ref{p1} the second assertion  follows from the first
one (since
$\sigma\textup{-}\!\operatorname{co}f\ge\mu\textup{-}\!\operatorname{co}f=\overline{\operatorname{co}}f$).
The lemma is proved.
\end{proof}

We will also use the following corollary of boundedness of the set
$\mathfrak{S}(\mathcal{H})$ as a subset of
$\mathfrak{T}(\mathcal{H})$.

\begin{lemma}
\label{l2} Let $f$ be a concave upper semicontinuous function on a
convex subset $\mathcal{A}\subseteq\mathfrak{S}(\mathcal{H})$. If
the function $f$ is finite at a particular state in $\mathcal{A}$
then this function is upper bounded on the set $\mathcal{A}$.
\end{lemma}

\begin{proof}
Let $\rho_{0}$ be such state in $\mathcal{A}$ that
$f(\rho_{0})=c_{0}\neq\pm\infty$. Without loss of generality we can
consider that $c_{0}=0$. If there exists a sequence
$\{\rho_{n}\}\subset\mathcal{A}$ such that
$\,\lim_{n\rightarrow+\infty}f(\rho_{n})=+\infty\,$ then the
sequence of states
$\sigma_{n}=(1-\lambda_{n})\rho_{0}+\lambda_{n}\rho_{n}$ in
$\mathcal{A}$, where $\lambda_{n}=(f(\rho_{n}))^{-1}$, converges to
the state $\rho_{0}$ by boundedness of the set $\mathcal{A}$ and
$f(\sigma_{n})\ge\lambda_{n}f(\rho_{n})=1$ by concavity of the
function $f$, contradicting to upper semicontinuity of this
function.
\end{proof}

\subsection{The convex roofs}
\label{s2.3} In the case $\dim\mathcal{H}<+\infty$ any state in
$\mathfrak{S}(\mathcal{H})$ can be represented as the average state
of some finite ensemble of pure states. This provides correctness of
the following convex extension to the set
$\mathfrak{S}(\mathcal{H})$ of an arbitrary function $f$ defined on
the set $\operatorname{extr}\mathfrak{S}(\mathcal{H})$ of pure
states
\begin{equation}
\label{eq8}
f_{*}(\rho)=\inf_{\{\pi_{i},\rho_{i}\}\in\widehat{\mathcal{P}}_{\{\rho\}}^{\mathrm{f}}}
\sum_{i}\pi_{i}f(\rho_{i}),
\qquad
\rho\in\mathfrak{S}(\mathcal{H})
\end{equation}
(the infimum is over all finite ensembles
$\{\pi_{i},\rho_{i}\}$ of \textit{pure} states with the average
state $\rho$). Following \cite{1} we will call this extension the
\textit{convex roof} of the function $f$. The notion of a convex roof
plays essential role in quantum information theory, where it is used,
in particular, for construction of entanglement monotones
(see~\S\,\ref{s4}).

In the case $\dim\mathcal{H}=+\infty$ one can consider the following
two generalizations of the above construction.

Let $f$ be a semibounded function $f$ on
the set $\operatorname{extr}\mathfrak{S}(\mathcal{H})$ of pure
states. The \textit{$\sigma$-convex roof} $f_{*}^{\sigma}$ of the function $f$ is defined as follows
\begin{equation}
\label{eq9}
f_{*}^{\sigma}(\rho)=\inf_{\{\pi_{i},\rho_{i}\}\in\widehat{\mathcal{P}}_{\{\rho\}}}
\sum_{i}\pi_{i}f(\rho_{i}),\qquad \rho\in\mathfrak{S}(\mathcal{H})
\end{equation}
(the infimum is over all countable ensembles
$\{\pi_{i},\rho_{i}\}$ of \textit{pure} states with the average
state $\rho$). Similar to the case of function
$\sigma\textup{-}\!\operatorname{co}f$ it is easy to show
$\sigma$-convexity of the function $f_{*}^{\sigma}$. Thus
$f_{*}^{\sigma}$ is the greatest $\sigma$-convex extension of the
function $f$ to the set $\mathfrak{S}(\mathcal{H})$.

Let $f$ be a semibounded Borel function $f$ on
the set $\operatorname{extr}\mathfrak{S}(\mathcal{H})$ of pure
states. The \textit{$\mu$-convex roof} $f_{*}^{\mu}$ of the function $f$ is defined as follows
\begin{equation}
\label{eq10}
f_{*}^{\mu}(\rho)=\inf_{\mu\in\widehat{\mathcal{P}}_{\{\rho\}}}
\int_{\operatorname{extr}\mathfrak{S}(\mathcal{H})}f(\sigma)\mu(d\sigma),\qquad
\rho\in\mathfrak{S}(\mathcal{H})
\end{equation}
(the infimum is over all probability measures $\mu$ supported
by \textit{pure} states  with the barycenter $\rho$). If the
function $f_{*}^{\mu}$ is universally measurable\footnote{By using
the results in \cite{19} universal measurability of the function
$f_{*}^{\mu}$ can be proved for any bounded Borel function $f$.} and
$\mu$-convex then it is the greatest $\mu$-convex extension of the
function $f$ to the set $\mathfrak{S}(\mathcal{H})$.  By
propositions \ref{p3} and \ref{p4} below (used with evident
convexity of the function $f_{*}^{\mu}$ and Proposition~\ref{pA-2}
in the Appendix) this holds if the function $f$ is either lower bounded
and lower semicontinuous or upper bounded and upper semicontinuous.

Note that the notions of a $\sigma$-convex roof and of a $\mu$-convex roof
can be reduced respectively to the notions of a $\sigma$-convex hull
and of a $\mu$-convex hull introduced in Section~\ref{s2.2}. Indeed, it
is easy to see that
$f_{*}^{\sigma}=\sigma\textup{-}\!\operatorname{co}\hat{f}$ and
$f_{*}^{\mu}=\mu\textup{-}\!\operatorname{co}\hat{f}$ for any
function $f$ on the set
$\operatorname{extr}\mathfrak{S}(\mathcal{H})$, where
$$
\hat{f}(\rho)=
\begin{cases}f(\rho),
&\rho\in \operatorname{extr}\mathfrak{S}(\mathcal{H}),
\\
+\infty,
&\rho\in \mathfrak{S}(\mathcal{H})\setminus\operatorname{extr}\mathfrak{S}(\mathcal{H}).
\end{cases}
$$
Since lower semicontinuity of the function  $f$ on the set
$\operatorname{extr}\mathfrak{S}(\mathcal{H})$ implies lower
semicontinuity of the function $\hat{f}$ on the set
$\mathfrak{S}(\mathcal{H})$, Proposition \ref{p1} implies the
following result (also derived from
assertion A of Theorem 2 in \cite{13} by means of $\mu$-compactness of the set
$\operatorname{extr}\mathfrak{S}(\mathcal{H})$).

\begin{proposition}
\label{p3} Let $f$ be a lower semicontinuous lower
bounded function on the set
$\,\operatorname{extr}\mathfrak{S}(\mathcal{H})$. Then
\begin{itemize}
    \item the function $f_{*}^{\mu}$ is the greatest lower semicontinuous
convex extension of the function $f$ to the set
$\,\mathfrak{S}(\mathcal{H})$;
    \item for arbitrary state $\rho$ in
$\,\mathfrak{S}(\mathcal{H})$ the infimum in the definition of the
value $f_{*}^{\mu}(\rho)$ in~\eqref{eq10} is achieved at some
measure in
$\,\widehat{\mathcal{P}}_{\{\rho\}}$.
\end{itemize}
\end{proposition}

Importance of the $\mu$-compactness property of the set
$\mathfrak{S}(\mathcal{H})$ in the proof of this proposition is
illustrated by the examples in \cite{15}.

By Theorem\,1 in \cite{13} stability of the set
$\mathfrak{S}(\mathcal{H})$ implies\footnote{By the generalized
Vesterstrom-O'Brien theorem proved in~\cite{15} openness of this map
is equivalent to stability of the set $\mathfrak{S}(\mathcal{H})$.}
openness of the map
$\mathcal{P}(\operatorname{extr}\mathfrak{S}(\mathcal{H}))\ni\mu\mapsto
\bar{\rho}(\mu)\in \mathfrak{S}(\mathcal{H})$. Hence assertion
 B of Theorem 2 in \cite{13} implies the following result.

\begin{proposition}
\label{p4} Let $f$ be an upper semicontinuous upper bounded function
on the set $\,\operatorname{extr}\mathfrak{S}(\mathcal{H})$. Then
\begin{itemize}
    \item the $\sigma$-convex roof and the $\mu$-convex roof
    of the function $f$ coincide: $f_{*}^{\sigma}=f_{*}^{\mu};$
    \item the function $f_{*}^{\sigma}=f_{*}^{\mu}$ is upper semicontinuous on the set
$\,\mathfrak{S}(\mathcal{H})$ and coincides with the greatest upper
bounded convex extension of the function $f$ to this set.
\end{itemize}
\end{proposition}

Propositions \ref{p3} and \ref{p4} have the following obvious
corollary.

\begin{corollary}
\label{c2} Let $f$ be a continuous bounded function on the set
$\,\operatorname{extr}\mathfrak{S}(\mathcal{H})$. Then its
$\sigma$-convex roof and its $\mu$-convex roof coincide and the
function $f_{*}^{\sigma}=f_{*}^{\mu}$ is continuous on the set
$\,\mathfrak{S}(\mathcal{H})$.
\end{corollary}

By this corollary an arbitrary continuous bounded function on the
set of pure states has at least one continuous bounded
convex extension to the set of all states.

\subsection{The convex hulls of concave functions}
\label{s2.4} In the case $\dim\mathcal{H}<+\infty$ it is easy to
show that the convex hull of arbitrary concave function $f$ defined
on the set $\mathfrak{S}(\mathcal{H})$ coincides with the convex
roof of the restriction
$f|_{\operatorname{extr}\mathfrak{S}(\mathcal{H})}$ of this function
to the set $\operatorname{extr}\mathfrak{S}(\mathcal{H})$. By
stability of the set $\mathfrak{S}(\mathcal{H})$
continuity of the function $f$ implies continuity of the function
$\operatorname{co}f=(f|_{\operatorname{extr}\mathfrak{S}(\mathcal{H})})_{*}$.

In the case $\dim\mathcal{H}=+\infty$ the analog of this
observation is established in the following proposition.

\begin{proposition}
\label{p5} Let $f$ be a concave  function on the set
$\mathfrak{S}(\mathcal{H})$.

If the function $f$ is lower bounded then
$\,\sigma\textup{-}\!\operatorname{co}f=\left(f|_{\operatorname{extr}\mathfrak{S}(\mathcal{H})}\right)_{*}^{\sigma}$.
If, in addition, the function $f$ is lower semicontinuous then
$\,\mu\textup{-}\!\operatorname{co}f=\left(f|_{\operatorname{extr}\mathfrak{S}(\mathcal{H})}\right)_{*}^{\mu}$
and this function is lower semicontinuous.

If the function $f$ is upper semicontinuous (correspondingly,
continuous and lower bounded)  then
$$
\operatorname{co}f=\sigma\textup{-}\!\operatorname{co}f=\mu\textup{-}\!\operatorname{co}f=
(f|_{\operatorname{extr}\mathfrak{S}(\mathcal{H})})_{*}^{\sigma}=
(f|_{\operatorname{extr}\mathfrak{S}(\mathcal{H})})_{*}^{\mu}
$$
and this function is upper semicontinuous (correspondingly,
continuous).
\end{proposition}

\begin{proof}
To show coincidence of the functions
$\sigma\textup{-}\!\operatorname{co}f$ and
$(f|_{\operatorname{extr}\mathfrak{S}(\mathcal{H})})_{*}^{\sigma}$
(correspondingly, of the functions
$\mu\textup{-}\!\operatorname{co}f$ and
$(f|_{\operatorname{extr}\mathfrak{S}(\mathcal{H})})_{*}^{\mu}$) it
is sufficient to prove the inequality
$\sigma\textup{-}\!\operatorname{co}f\ge(f|_{\operatorname{extr}\mathfrak{S}(\mathcal{H})}
)_{*}^{\sigma}$ (correspondingly, the inequality
$\mu\textup{-}\!\operatorname{co}f\ge(f|_{\operatorname{extr}\mathfrak{S}(\mathcal{H})})_{*}^{\mu}$).

The first inequality for the concave lower bounded function $f$
directly follows from the discrete Jensen's inequality
(Proposition~\ref{pA-1} in the Appendix).

Let $f$ be a lower bounded lower semicontinuous concave function and
$\rho_{0}$ be an arbitrary state. By Lemma \ref{l1} there exists
a sequence $\{\rho_{n}\}$, converging to the state $\rho_{0}$, such
that
$\lim_{n\rightarrow+\infty}\sigma\textup{-}\!\operatorname{co}f(\rho_{n})
=\mu\textup{-}\!\operatorname{co}f(\rho_{0})$. By the proved part of
the proposition we have
$$
\sigma\textup{-}\!\operatorname{co}f(\rho_{n})
=(f|_{\operatorname{extr}\mathfrak{S}(\mathcal{H})})_{*}^{\sigma}(\rho_{n})
\ge(f|_{\operatorname{extr}\mathfrak{S}(\mathcal{H})})_{*}^{\mu}(\rho_{n})
\qquad \forall\, n.
$$
By Proposition \ref{p3}  passing to the limit
$n\rightarrow+\infty$ in this inequality leads to the
inequality
$\mu\textup{-}\!\operatorname{co}f
(\rho_{0})\ge(f|_{\operatorname{extr}\mathfrak{S}(\mathcal{H})})_{*}^{\mu}(\rho_{0})$.

Let $f$ be an upper semicontinuous concave function taking finite value at least at one state. By Lemma \ref{l2} this function is upper
bounded. Propositions \ref{p2} and  \ref{p4} imply respectively
$\operatorname{co}f=\sigma\textup{-}\!\operatorname{co}f=\mu\textup{-}\!\operatorname{co}f$
and
$(f|_{\operatorname{extr}\mathfrak{S}(\mathcal{H})})_{*}^{\sigma}=
(f|_{\operatorname{extr}\mathfrak{S}(\mathcal{H})})_{*}^{\mu}$ as
well as upper semicontinuity of these functions. Since
$\operatorname{co}f\ge(f|_{\operatorname{extr}\mathfrak{S}(\mathcal{H})})_{*}^{\sigma}$
by Proposition~\ref{pA-2} in the Appendix and
$\mu\textup{-}\!\operatorname{co}f\le(f|_{\operatorname{extr}\mathfrak{S}(\mathcal{H})})_{*}^{\mu}$
by the definitions, we obtain the main part of the second assertion of the
proposition.

The assertion concerning concave continuous lower bounded function
$f$ follows from the previous ones.
\end{proof}

\subsection{One result concerning the convex closure}
\label{s2.5} It is well known\footnote{It follows from Dini's lemma.
The importance of the compactness condition can be shown by the
sequence of the functions $f_{n}(x)=\exp({-x^{2}/n})$ on
$\mathbb{R}$, converging to the function $f_{0}(x)\equiv 1$, such
that $\overline{\operatorname{co}}f_{n}(x)\equiv 0$ for all $n$.}
that for an arbitrary increasing sequence $\{f_{n}\}$ of continuous
functions on a convex compact set $\mathcal{A}$, pointwise
converging to a continuous function $f_{0}$, the corresponding
sequence $\{\overline{\operatorname{co}}f_{n}\}$ converges to the
function $\overline{\operatorname{co}}f_{0}$. It turns out that the
$\mu$-compactness of the set $\mathfrak{S}(\mathcal{H})$ implies the
analogous observation.
\begin{proposition}
\label{p6} For arbitrary increasing sequence $\{f_{n}\}$ of lower
semicontinuous lower bounded functions on the set
$\,\mathfrak{S}(\mathcal{H})$ and arbitrary converging sequence
$\{\rho_{n}\}$ of states in $\,\mathfrak{S}(\mathcal{H})$ the
following inequality holds
$$
\liminf_{n\rightarrow+\infty}\overline{\operatorname{co}}f_{n}(\rho_{n})
\ge\overline{\operatorname{co}}f_{0}(\rho_{0}),
$$
where $f_{0}=\sup_{n}f_{n}$ and
$\rho_{0}=\lim_{n\rightarrow+\infty}\rho_{n}$. In particular,
$$
\lim_{n\rightarrow+\infty}\overline{\operatorname{co}}f_{n}(\rho)
=\overline{\operatorname{co}}f_{0}(\rho)
\qquad
\forall\, \rho\in\mathfrak{S}(\mathcal{H}).
$$
\end{proposition}

In fact, $\mu$-compactness of the set $\mathfrak{S}(\mathcal{H})$ is
\textit{equivalent} to validity of the last assertion of Proposition
\ref{p6} (see~\cite{23}).

\begin{proof}
For an arbitrary Borel function $g$ on the set
$\mathfrak{S}(\mathcal{H})$ and any measure
$\mu\in\mathcal{P}$ we will use the following notation:
$$
\mu(g)=\int_{\mathfrak{S}(\mathcal{H})}g(\sigma)\mu(d\sigma).
$$

Without loss of generality we may assume that the sequence
$\{f_{n}\}$ consists of nonnegative functions. Suppose there exists
a sequence $\{\rho_{n}\}$, converging to a state $\rho_{0}$,
such that
$$
\overline{\operatorname{co}}f_{n}(\rho_{n})+\Delta\le\overline{\operatorname{co}}f_{0}(\rho_{0}),
\qquad \Delta>0, \qquad \forall\, n.
$$
We will assume that
$\overline{\operatorname{co}}f_{0}(\rho_{0})<+\infty$. The case
$\overline{\operatorname{co}}f_{0}(\rho_{0})=+\infty$ is considered
similarly.

By representation \eqref{eq4} there exists a continuous affine
function $\alpha$ on the set $\mathfrak{S}(\mathcal{H})$ such that
\begin{equation}
\label{eq11}
\alpha(\rho)\le f_{0}(\rho)\quad\forall\,
\rho\in\mathfrak{S}(\mathcal{H}),
\qquad
\overline{\operatorname{co}}f_{0}(\rho_{0})\le\alpha(\rho_{0})+\frac{1}{4}\Delta.
\end{equation}

Let $N$ be such number that
$|\alpha(\rho_{n})-\alpha(\rho_{0})|<\frac{1}{4}\Delta$ for all
$n\ge N$.

By Proposition \ref{p1} for each $n$ there exists a measure
$\mu_{n}\in\mathcal{P}_{\{\rho_{n}\}}$ such that
$\overline{\operatorname{co}}f_{n}(\rho_{n})=\mu_{n}(f_{n})$. Since
the function $\alpha$ is affine we have
\begin{align}
\nonumber
\mu_{n}(\alpha)-\mu_{n}(f_{n})
&=\alpha(\rho_{n})-\overline{\operatorname{co}}f_{n}(\rho_{n})
\\
\nonumber
&=[\alpha(\rho_{n})-\alpha(\rho_{0})]+[\alpha(\rho_{0})-\overline{\operatorname{co}}f_{0}(\rho_{0})]
+[\overline{\operatorname{co}}f_{0}(\rho_{0})-\overline{\operatorname{co}}f_{n}(\rho_{n})\,]
\\
&\ge-\frac{1}{4}\Delta-\frac{1}{4}\Delta+\Delta=\frac{1}{2}\Delta
\qquad\forall\,n\ge N.
\label{eq12}
\end{align}

The $\mu$-compactness of the set $\mathfrak{S}(\mathcal{H})$ implies
relative compactness of the sequence $\{\mu_{n}\}$. By
Prokhorov's theorem (see~\cite{24}, \S\,6) this sequence is
\textit{tight}, which means existence of such compact subset
$\mathcal{K}_{\varepsilon}\subset\mathfrak{S}(\mathcal{H})$ for each
$\varepsilon>0$ that
$\mu_{n}(\mathfrak{S}(\mathcal{H})\setminus\mathcal{K}_{\varepsilon})<\varepsilon$
for all~$n$.

Let $M=\sup_{\rho\in\mathfrak{S}(\mathcal{H})}|\alpha(\rho)|$ and
$\varepsilon_{0}=\frac{\Delta}{4M}$. By \eqref{eq12} for all  $n\ge
N$ we have
$$
\int_{\mathcal{K}_{\varepsilon_{0}}}(\alpha(\rho)-f_{n}(\rho))\mu_{n}(d\rho)
\ge\frac{1}{2}\Delta-\int_{\mathfrak{S}(\mathcal{H})\setminus\mathcal{K}_{\varepsilon_{0}}}
(\alpha(\rho)-f_{n}(\rho))\mu_{n}(d\rho)
\ge \frac{1}{4}\Delta.
$$
Hence, the set
$\mathcal{C}_{n}=\{\rho\in\mathcal{K}_{\varepsilon_{0}}\mid
\alpha(\rho)\ge f_{n}(\rho)+\frac{1}{4}\Delta\}$ is nonempty for all
$n\ge N$.

Since the sequence $\{f_{n}\}$ is increasing, the sequence
$\{\mathcal{C}_{n}\}$ of \textit{closed} subsets of the
\textit{compact} set $\mathcal{K}_{\varepsilon_{0}}$ is monotone:
$\mathcal{C}_{n+1}\subseteq\mathcal{C}_{n}$\enskip $\forall\, n$.
Hence there exists $\rho_{*}\in\bigcap_{n}\mathcal{C}_{n}$. This
means that $\alpha(\rho_{*})\geq f_{n}(\rho_{*})+\frac{1}{4}\Delta$
for all $n$, and hence $\alpha(\rho_{*})> f_{0}(\rho_{*})$,
contradicting to~\eqref{eq11}.
\end{proof}

\begin{corollary}
\label{c3} For arbitrary increasing sequence $\{f_{n}\}$ of lower
semicontinuous lower bounded functions on the set
$\,\operatorname{extr}\mathfrak{S}(\mathcal{H})$ and arbitrary
converging sequence $\{\rho_{n}\}$ of states in
$\,\mathfrak{S}(\mathcal{H})$ the following inequality holds
$$
\liminf_{n\rightarrow+\infty}(f_{n})_{*}^{\mu}(\rho_{n})\ge(f_{0})_{*}^{\mu}(\rho_{0}),
$$
where $f_{0}=\sup_{n}f_{n}$ and
$\rho_{0}=\lim_{n\rightarrow+\infty}\rho_{n}$. In particular,
$$
\lim_{n\rightarrow+\infty}(f_{n})_{*}^{\mu}(\rho)=(f_{0})_{*}^{\mu}(\rho)
\qquad
\forall\, \rho\in\mathfrak{S}(\mathcal{H}).
$$
\end{corollary}

\begin{proof} By Theorems 1 and 2 in \cite{13} for an arbitrary lower
semicontinuous lower bounded function $f$ on the set
$\operatorname{extr}\mathfrak{S}(\mathcal{H})$ the function
$$
f^{*}(\rho)\doteq\sup_{\mu\in\widehat{\mathcal{P}}_{\{\rho\}}}
\int_{\operatorname{extr}\mathfrak{S}(\mathcal{H})}f(\sigma)\mu(d\sigma)
=\sup_{\{\pi_{i},\rho_{i}\}\in\widehat{\mathcal{P}}_{\{\rho\}}}
\sum_{i}\pi_{i}f(\rho_{i}),\qquad \rho\in\mathfrak{S}(\mathcal{H}),
$$
is a lower semicontinuous lower bounded concave extension of the
function $f$ to the set $\mathfrak{S}(\mathcal{H})$. It is clear
that for an arbitrary increasing sequence $\{f_{n}\}$ of lower
semicontinuous lower bounded functions on the set
$\operatorname{extr}\mathfrak{S}(\mathcal{H})$, converging pointwise to the
function $f_{0}$, the corresponding increasing sequence
$\{f_{n}^{*}\}$ converges pointwise to the function $f_{0}^{*}$ on the set
$\mathfrak{S}(\mathcal{H})$. Thus the assertion of the corollary can
be derived from Proposition \ref{p6} by using Propositions \ref{p1}
and \ref{p5}.
\end{proof}

\begin{remark}
\label{r1} The $\mu$-convex roof can not be replaced by the
$\sigma$-convex roof in Corollary\,\ref{c3}. Indeed, let $f$ be the
indicator function of the set
$\operatorname{extr}\mathfrak{S}(\mathcal{H}\otimes\mathcal{H})\setminus\mathcal{A}_{s}$
and $\omega_{0}$~be the separable state considered in Example
\ref{e2}. This function $f$ can be represented as a limit of an
increasing sequence $\{f_{n}\}$ of continuous bounded functions on
the set
$\operatorname{extr}\mathfrak{S}(\mathcal{H}\otimes\mathcal{H})$.
Since by Corollary \ref{c2} we have
$(f_{n})_{*}^{\sigma}=(f_{n})_{*}^{\mu}$ for all $n$, Corollary\,\ref{c3} and the property of the state $\omega_{0}$ imply
$\lim_{n\rightarrow+\infty}(f_{n})_{*}^{\sigma}(\omega_{0})=(f_{0})_{*}^{\mu}(\omega_{0})=0$
and $(f_{0})_{*}^{\sigma}(\omega_{0})=1$.
\end{remark}

\begin{remark}
\label{r2} The monotone convergence theorem implies the following
results dual to the second assertions of Proposition \ref{p6} and of
Corollary \ref{c3}:

1) \textit{For an arbitrary decreasing sequence $\,\{f_{n}\}$ of Borel upper bounded functions on the set
$\,\mathfrak{S}(\mathcal{H})$ the following relation holds}
$$
\lim_{n\rightarrow+\infty}\mu\textup{-}\!\operatorname{co}f_{n}(\rho)
=\mu\textup{-}\!\operatorname{co}f_{0}(\rho),\quad \forall\,
\rho\in\mathfrak{S}(\mathcal{H}), \qquad \textit{where}\quad
f_{0}=\inf_{n}f_{n};
$$

2) \textit{For an arbitrary decreasing sequence $\,\{f_{n}\}$ of Borel
upper bounded functions on the set
$\,\operatorname{extr}\mathfrak{S}(\mathcal{H})$ the following
relation holds
$$
\lim_{n\rightarrow+\infty}(f_{n})_{*}^{\mu}(\rho)=(f_{0})_{*}^{\mu}(\rho),\qquad
\forall\, \rho\in\mathfrak{S}(\mathcal{H}), \qquad
\textit{where}\quad f_{0}=\inf_{n}f_{n}.
$$}
\end{remark}

By using Corollary \ref{c1}, Proposition \ref{p6}, the first
assertion of Remark \ref{r2} and Dini's lemma the following result
can be easily proved.

\begin{corollary}
\label{c4} Let $\{f_{t}\}_{t\in\mathrm{T}\subseteq\mathbb{R}}$~be a
family of continuous bounded functions on the set
$\,\mathfrak{S}(\mathcal{H})$ such that:

1) $f_{t_{1}}(\rho)\le f_{t_{2}}(\rho)$ for all
$\rho\in\mathfrak{S}(\mathcal{H})$ and all
$t_{1},t_{2}\in\mathrm{T}$ such that $t_{1}<t_{2}$;

2) the function  $\mathrm{T}\ni t\mapsto f_{t}(\rho)$ is continuous
for all $\rho\in\mathfrak{S}(\mathcal{H})$.

Then the function $\mathfrak{S}(\mathcal{H})\times\mathrm{T}\ni
(\rho,t)\mapsto \operatorname{co}f_{t}(\rho)$ is continuous.
\end{corollary}

By using Corollary \ref{c2}, Corollary \ref{c3}, the second
assertion of Remark \ref{r2} and Dini's lemma the analogous result
for the $\mu$-convex roof of a family of continuous
bounded functions on the set
$\operatorname{extr}\mathfrak{S}(\mathcal{H})$ can be proved.

\section{The main theorem}
\label{s3}

Let $\alpha$ be a lower semicontinuous affine function on the set
$\mathfrak{S}(\mathcal{H})$ taking values in $[0, +\infty]$.
Consider the family of closed subsets
\begin{equation}
\label{eq13}
\mathcal{A}_{c}=\{\rho\in\mathfrak{S}(\mathcal{H})\mid \alpha(\rho)\le c\},
\qquad c\in \mathbb{R}_{+},
\end{equation}
of the set  $\mathfrak{S}(\mathcal{H})$. In the following theorem
the properties of restrictions of convex hulls of a given function to the subsets of
this family are considered.

\begin{theorem}
\label{t1} Let $f$ be a Borel lower bounded function on the set
$\,\mathfrak{S}(\mathcal{H})$ and $\alpha$ be the above affine
function. If the function $f$ has upper semicontinuous bounded
restriction to the set $\mathcal{A}_{c}$ for each $c>0$ and
\begin{equation}
\label{eq14}
\limsup_{c\rightarrow+\infty}\;c^{-1}\sup_{\rho\in\mathcal{A}_{c}}
f(\rho)<+\infty,
\end{equation}
then
$$
\operatorname{co}f(\rho)=\sigma\textup{-}\!\operatorname{co}f(\rho)=\mu\textup{-}\!\operatorname{co}f(\rho)
$$
for all $\rho\in\bigcup_{c>0}\mathcal{A}_{c}$ and the common
restriction of these functions to the set $\mathcal{A}_{c}$  is
upper semicontinuous for each $c>0$.

If, in addition, the function $f$ is lower semicontinuous on the set
$\,\mathfrak{S}(\mathcal{H})$ then
$$
\operatorname{co}f(\rho)=\sigma\textup{-}\!\operatorname{co}f(\rho)
=\mu\textup{-}\!\operatorname{co}f(\rho)=\overline{\operatorname{co}}f(\rho)
$$
for all $\rho\in\bigcup_{c>0}\mathcal{A}_{c}$ and the common
restriction of these functions to the set $\mathcal{A}_{c}$ is
continuous for each $c>0$.
\end{theorem}

\begin{proof}
Without loss of generality we can assume that $f$ is a nonnegative
function.

Let $\rho_{0}$ be a state such that
$\alpha(\rho_{0})=c_{0}<+\infty$. By the condition
$\mu\textup{-}\!\operatorname{co}f(\rho_{0})\le
f(\rho_{0})<+\infty$. For arbitrary $\varepsilon>0$ let $\mu_{0}$~be
a measure in $\mathcal{P}_{\{\rho_{0}\}}$ such that
$$
\int_{\mathfrak{S}(\mathcal{H})}
f(\rho)\mu_{0}(d\rho)<\mu\textup{-}\!\operatorname{co}f(\rho_{0})+\varepsilon.
$$
Condition \eqref{eq14} implies  existence of such positive numbers
$c_{*}$ and $M$ that $f(\rho)\le M\alpha(\rho)$ for all
$\rho\in\mathfrak{S}(\mathcal{H})\setminus\mathcal{A}_{c_{*}}$.

Note that $\lim_{c\rightarrow+\infty}\mu_{0}(\mathcal{A}_{c})=1$.
Indeed, it follows from the inequality
$$
c\mu_{0}(\mathfrak{S}(\mathcal{H})\setminus\mathcal{A}_{c})
\le\int_{\mathcal{A}_{c}}\alpha(\rho)\mu_{0}(d\rho)+
\int_{\mathfrak{S}(\mathcal{H})\setminus\mathcal{A}_{c}}\alpha(\rho)\mu_{0}(d\rho)=
\alpha(\rho_{0})=c_{0},
$$
obtained by using Corollary \ref{cA-1} in the Appendix that
$$
\mu_{0}(\mathfrak{S}(\mathcal{H})\setminus\mathcal{A}_{c})\le
\frac{c_{0}}{c}.
$$
Thus the monotone convergence theorem implies
$$
\lim_{c\rightarrow+\infty}\int_{\mathfrak{S}(\mathcal{H})
\setminus\mathcal{A}_{c}}\alpha(\rho)\mu_{0}(d\rho)=
\lim_{c\rightarrow+\infty}\biggl(\alpha(\rho_{0})-\int_{\mathcal{A}_{c}}\alpha(\rho)\mu_{0}(d\rho)\biggr)=0.
$$
Let $c^{*}>c_{*}$ be such that
$\int_{\mathfrak{S}(\mathcal{H})\setminus\mathcal{A}_{c^{*}}}\alpha(\rho)\mu_{0}(d\rho)<\varepsilon$.
By Lemma \ref{l3} below there exists a sequence $\{\mu_{n}\}$ of
measures in $\mathcal{P}_{\{\rho_{0}\}}^{\mathrm{f}}$ weakly
converging to the measure $\mu_{0}$ such that
$\mu_{n}(\mathcal{A}_{c^{*}})=\mu_{0}(\mathcal{A}_{c^{*}})$ and
$\int_{\mathfrak{S}(\mathcal{H})\setminus\mathcal{A}_{c^{*}}}\alpha(\rho)\mu_{n}(d\rho)<\varepsilon$
for all $n$. Since the function $f$ is upper semicontinuous and
bounded on the set $\mathcal{A}_{c^{*}}$ we have
(see~\cite{24}, \S\,2)
$$
\limsup_{n\rightarrow+\infty}\int_{\mathcal{A}_{c^{*}}}f(\rho)\mu_{n}(d\rho)
\le\int_{\mathcal{A}_{c^{*}}}f(\rho)\mu_{0}(d\rho).
$$
Hence by noting that
$$
\int_{\mathfrak{S}(\mathcal{H})\setminus\mathcal{A}_{c^{*}}}f(\rho)\mu_{n}(d\rho)\le
M\int_{\mathfrak{S}(\mathcal{H})\setminus\mathcal{A}_{c^{*}}}\alpha(\rho)\mu_{n}(d\rho)<M\varepsilon,\qquad
n=0,1,2,\dots,
$$
we obtain
\begin{align*}
\operatorname{co}f(\rho_{0})
&\le
\liminf_{n\rightarrow+\infty}\int_{\mathfrak{S}(\mathcal{H})}f(\rho)\mu_{n}(d\rho)\le
\limsup_{n\rightarrow+\infty}\int_{\mathcal{A}_{c^{*}}}f(\rho)\mu_{n}(d\rho)+M\varepsilon
\\
&\le
\int_{\mathfrak{S}(\mathcal{H})}f(\rho)\mu_{0}(d\rho)+M\varepsilon\le
\mu\textup{-}\!\operatorname{co}f(\rho_{0})+\varepsilon(M+1).
\end{align*}
Since $\varepsilon$ is arbitrary this implies
$\operatorname{co}f(\rho_{0})=\mu\textup{-}\!\operatorname{co}f(\rho_{0})$.

The proof of the first assertion of the theorem is completed by
applying Lemma\,\ref{l4} below.

By Proposition \ref{p1} the second assertion of the theorem follows
from the first one.
\end{proof}

\begin{lemma}
\label{l3} Let $\alpha$ be a lower semicontinuous affine function on
the set $\,\mathfrak{S}(\mathcal{H})$ taking values in $\,[0,+\infty]$
and $\mu_{0}$ be a measure in $\mathcal{P}$ such that
$\alpha(\bar{\rho}(\mu_{0}))<+\infty$. For given arbitrary
$c\,{>}\,0$ there exists a sequence $\{\mu_{n}\}$ of measures in
$\mathcal{P}^{\mathrm{f}}_{\{\bar{\rho}(\mu_{0})\}}$ converging to
the measure $\mu_{0}$ such that
$$
\mu_{n}(\mathcal{A}_{c})=\mu_{0}(\mathcal{A}_{c}),
\qquad
\int_{\mathfrak{S}(\mathcal{H})\setminus\mathcal{A}_{c}}\alpha(\rho)\mu_{n}(d\rho)=
\int_{\mathfrak{S}(\mathcal{H})\setminus\mathcal{A}_{c}}\alpha(\rho)\mu_{0}(d\rho)
$$
for all $n$, where $\mathcal{A}_{c}$~is the subset of
$\mathfrak{S}(\mathcal{H})$ defined by~\eqref{eq13}.
\end{lemma}

\begin{proof}
This lemma can be proved by the simple modification of the proof of
Lemma\,1 in ~\cite{4}, consisting in finding for given $n$ of such
decomposition of the set $\mathfrak{S}(\mathcal{H})$ into collection
$\{\mathcal{A}_{i}^{n}\}_{i=1}^{m+2}$ of $m+2$\enskip ($m=m(n)$)
disjoint Borel subsets that

1) the set $\mathcal{A}_{i}^{n}$ has diameter $<1/n$
  for $i=\overline{1,m}$;

2) $\mu_{0}(\mathcal{A}_{m+1}^{n})<1/n$ and
$\mu_{0}(\mathcal{A}_{m+2}^{n})<1/n$;

3) the set $\mathcal{A}_{i}^{n}$ is contained either in
$\mathcal{A}_{c}$ or in
$\mathfrak{S}(\mathcal{H})\setminus\mathcal{A}_{c}$ for
$i=\overline{1,m+2}$.

The essential points in this construction are the following
implication
$$
\mathcal{A}\subseteq\mathcal{B}
\ \Rightarrow\
(\mu_{0}(\mathcal{A}))^{-1}\int_{\mathcal{A}}\rho\mu_{0}(d\rho)\in\mathcal{B},
\qquad
\mathcal{B}=\mathcal{A}_{c},
\quad
\mathfrak{S}(\mathcal{H})\setminus\mathcal{A}_{c},
$$
and the equality
$$
\int_{\mathcal{A}}\alpha(\rho)\mu_{0}(d\rho)=
\mu_{0}(\mathcal{A})\,\alpha\biggl(\frac{1}{\mu_{0}(\mathcal{A})}
\int_{\mathcal{A}}\rho\mu_{0}(d\rho)\biggr),
\qquad
\mathcal{A}\subseteq\mathfrak{S}(\mathcal{H}),
\quad
\mu_{0}(\mathcal{A})\neq
0,
$$
easily obtained by using Corollary~\ref{cA-1} in the Appendix. The
lemma is proved.
\end{proof}

Stability of the set $\mathfrak{S}(\mathcal{H})$ is used in the
proof of the above Theorem \ref{t1} via the following lemma.

\begin{lemma}
\label{l4} Let $\alpha$ be a lower semicontinuous affine function on
the set $\,\mathfrak{S}(\mathcal{H})$ taking values in
$\,[0,+\infty]$ and $f$ be a function on the set
$\,\mathfrak{S}(\mathcal{H})$ having upper semicontinuous
restriction to the set $\mathcal{A}_{c}$ defined by~\eqref{eq13}
for each $c>0$. Then the function $\operatorname{co}f$ has upper
semicontinuous restriction to the set $\mathcal{A}_{c}$ for each
$c>0$.
\end{lemma}

\begin{proof}
Let $\rho_{0}\in\mathcal{A}_{c_{0}}$ and let
$\{\rho_{n}\}\subset\mathcal{A}_{c_{0}}$ be an arbitrary sequence
converging to the state $\rho_{0}$. Suppose there exists
\begin{equation}
\label{eq15}
\lim_{n\rightarrow+\infty}\operatorname{co}f(\rho_{n})>\operatorname{co}f(\rho_{0}).
\end{equation}

For given arbitrary $\varepsilon>0$ let $\{\pi_{i}^{0},
\rho_{i}^{0}\}_{i=1}^{m}$ be an ensemble in
$\mathcal{P}_{\{\rho_{0}\}}^{\mathrm{f}}$ such that
$\sum_{i=1}^{m}\pi_{i}^{0}f(\rho_{i}^{0})<\operatorname{co}f(\rho_{0})+\varepsilon$.
By stability of the set $\mathfrak{S}(\mathcal{H})$ (see~\cite{5})
there exists a sequence $\{\{\pi_{i}^{n},
\rho_{i}^{n}\}_{i=1}^{m}\}_{n}$ of ensembles such that
$\sum_{i=1}^{m}\pi_{i}^{n}\rho_{i}^{n}=\rho_{n}$ for each $n$,
$\lim_{n\rightarrow+\infty}\pi_{i}^{n}=\pi_{i}^{0}$ and
$\lim_{n\rightarrow+\infty}\rho_{i}^{n}=\rho_{i}^{0}$ for all
$i=\overline{1,m}$. Let $\pi_{*}=\min_{1\le i\le m}\pi_{i}^{0}$.
Then there exists such $N$ that $\pi_{i}^{n}\ge \pi_{*}/2$ for all
$n\ge N$ and $i=\overline{1,m}$. It follows from the inequality
$\sum_{i=1}^{m}\pi_{i}^{n}\alpha(\rho_{i}^{n})=\alpha(\rho_{n})\le
c_{0}$ that $\rho_{i}^{n}\in \mathcal{A}_{2c_{0}/\pi_{*}}$ for all
$n\ge N$ and $i=\overline{1,m}$. By upper semicontinuity of the
function $f$ on the set $\mathcal{A}_{\frac{2c_{0}}{\pi_{*}}}$ we
have
$$ \limsup_{n\rightarrow+\infty}\operatorname{co}f(\rho_{n})
\le\limsup_{n\rightarrow+\infty}
\sum_{i=1}^{m}\pi_{i}^{n}f(\rho_{i}^{n})
\le\sum_{i=1}^{m}\pi_{i}^{0}f(\rho_{i}^{0})
<\operatorname{co}f(\rho_{0})+\varepsilon,
$$
which contradicts to \eqref{eq15} since $\varepsilon$ is arbitrary.
\end{proof}

\begin{remark}
\label{r3} If $f$ is a concave function then condition \eqref{eq14}
follows from boundedness of the restriction of this function to the
set $\mathcal{A}_{c}$ for each $c$. Indeed, for arbitrary affine
function $\alpha$ concavity of the function $f$ on the set
$\mathfrak{S}(\mathcal{H})$ implies concavity of the function
$c\mapsto\sup_{\rho\in\mathcal{A}_{c}}f(\rho)$ on the set
$\mathbb{R}_{+}$, hence finiteness of the last function guarantees validity of
condition \eqref{eq14}.
\end{remark}

By Remark~\ref{r3} Theorem \ref{t1}, Lemma \ref{l2} and Proposition \ref{p5}
imply the following result.

\begin{corollary}
\label{c5} Let $f$ be a concave lower semicontinuous lower bounded
function and $\alpha$ be a lower semicontinuous affine function on
the set $\,\mathfrak{S}(\mathcal{H})$ taking values
in~$\,[0,+\infty]$. If the function $f$ has continuous restriction
to the set $\mathcal{A}_{c}$ defined by~\eqref{eq13} for each
$c>0$ then
$$
\operatorname{co}f(\rho)=\sigma\textup{-}\!\operatorname{co}f(\rho)=\mu\textup{-}\!\operatorname{co}f(\rho)=\overline{\operatorname{co}}f(\rho)=
(f|_{\operatorname{extr}\mathfrak{S}(\mathcal{H})})_{*}^{\sigma}(\rho)=
(f|_{\operatorname{extr}\mathfrak{S}(\mathcal{H})})_{*}^{\mu}(\rho)
$$
for all $\rho\in\bigcup_{c>0}\mathcal{A}_{c}$ and the common
restriction of these functions to the set $\mathcal{A}_{c}$ is
continuous for each $c>0$.
\end{corollary}

Theorem  \ref{t1} implies the following sufficient conditions of
coincidence and continuity of convex hulls.

\begin{corollary}
\label{c6} Let $f$ be a Borel lower bounded function on the set
$\,\mathfrak{S}(\mathcal{H})$ and $\rho_{0}$ be an arbitrary state
in $\,\mathfrak{S}(\mathcal{H})$. If there exists an affine lower
semicontinuous function $\alpha$ on the set
$\,\mathfrak{S}(\mathcal{H})$ taking values in~$[0, +\infty]$ such
that $\alpha(\rho_{0})<+\infty$, the function $f$ has upper
semicontinuous bounded restriction to the set $\mathcal{A}_{c}$
defined by~\eqref{eq13} for each $c>0$ and condition~\eqref{eq14}
holds, then
$$
\operatorname{co}f(\rho_{0})=\sigma\textup{-}\!\operatorname{co}f(\rho_{0})
=\mu\textup{-}\!\operatorname{co}f(\rho_{0}).
$$
\end{corollary}

\begin{corollary}
\label{c7} Let $f$ be a lower semicontinuous lower bounded function
on the set $\,\mathfrak{S}(\mathcal{H})$ and $\{\rho_{n}\}$ be an
arbitrary sequence of states in $\,\mathfrak{S}(\mathcal{H})$
converging to a state $\rho_{0}$. If there exists an affine lower
semicontinuous function $\alpha$ on the set
$\,\mathfrak{S}(\mathcal{H})$ taking values in $\,[0,+\infty]$ such
that $\,\sup_{n}\alpha(\rho_{n})<+\infty$, the function $f$ has
continuous bounded restriction to the set $\mathcal{A}_{c}$ defined
by~\eqref{eq13} for each $c>0$ and condition \eqref{eq14} holds,
then
\begin{gather}
\label{eq16}
\operatorname{co}f(\rho_{n})=\sigma\textup{-}\!\operatorname{co}f(\rho_{n})
=\mu\textup{-}\!\operatorname{co}f(\rho_{n})=\overline{\operatorname{co}}f(\rho_{n}),
\qquad
n=0,1,2,\dots,
\\
\label{eq17}
\lim_{n\rightarrow+\infty}\operatorname{co}f(\rho_{n})=\operatorname{co}f(\rho_{0}).
\end{gather}
\end{corollary}

\begin{remark}
\label{r4} If $f$ is a concave function then condition~\eqref{eq14}
in Corollaries \ref{c6} and \ref{c7} can be omitted by Remark \ref{r3}.
\end{remark}

\begin{example}
\label{e4} In study of informational properties of a quantum channel
the output Renyi entropy, in particular, the output von Neumann entropy and their
convex closures play important role \cite{25}.

Let
$\Phi\colon\mathfrak{T}(\mathcal{H})\mapsto\mathfrak{T}(\mathcal{H}')$~be
a quantum channel -- a linear completely positive
trace preserving map (see~\cite{9}, \S\,3.1) and
$\mathfrak{S}(\mathcal{H})\ni\rho\mapsto(R_{p}\circ\Phi)(\rho)
=\frac{\log\operatorname{Tr}\Phi(\rho)^{p}}{1-p}$~be the output
Renyi entropy of this channel of order $p\in(0,+\infty]$ (the case
$p=1$ corresponds to the output von Neumann entropy
$-\operatorname{Tr}\Phi(\rho)\log\Phi(\rho)$,  the case $p=+\infty$
corresponds to the function
$-\log\lambda_{\mathrm{max}}(\Phi(\rho))$, where
$\lambda_{\mathrm{max}}(\Phi(\rho))$~is the maximal eigenvalue of
the state $\Phi(\rho)$). For $p\in(0,1]$ the function
$R_{p}\circ\Phi$ is lower semicontinuous concave and takes values in
$[0,+\infty]$, while for  $p\in(1,+\infty]$ it is continuous and
finite but not concave. The output von Neumann entropy
$H\circ\Phi=R_{1}\circ\Phi$ is the supremum (pointwise limit as
$p\rightarrow 1+0$) of the monotonic family
$\{R_{p}\circ\Phi\}_{p>1}$ of continuous functions. By Proposition \ref{p6} the convex
closure $\overline{\operatorname{co}}(H\circ\Phi)$ of the output von
Neumann entropy coincides with the supremum (pointwise limit as
$p\rightarrow 1+0$) of the monotonic family of functions
$\{\overline{\operatorname{co}}(R_{p}\circ\Phi)\}_{p>1}$.

Corollary \ref{c6} makes it possible to show that
\begin{equation}
\label{eq18}
\begin{gathered}
\operatorname{co}(R_{p}\circ\Phi)(\rho_{0})
=\sigma\textup{-}\!\operatorname{co}(R_{p}\circ\Phi)(\rho_{0})
=\mu\textup{-}\!\operatorname{co}(R_{p}\circ\Phi)(\rho_{0})=
\overline{\operatorname{co}}(R_{p}\circ\Phi)(\rho_{0})
\\
\forall\, p\in[1,+\infty]
\end{gathered}
\end{equation}
for any state $\rho_{0}$ such that $(H\circ\Phi)(\rho_{0})<+\infty$. Indeed,  the condition
$H(\Phi(\rho_{0}))<+\infty$ implies existence of such
$\mathfrak{H}$-operator $H'$ in the space $\mathcal{H}'$ that
$$
\operatorname{g}(H')=\inf\{\lambda>0\mid \operatorname{Tr}\exp(-\lambda H')<+\infty\}<+\infty
$$
and $\operatorname{Tr}H'\Phi(\rho_{0})<+\infty$. By
Proposition 1 in \cite{26} the conditions of Corollary \ref{c6} are
fulfilled for the function
$f(\rho)=(R_{p}\circ\Phi)(\rho)\le(H\circ\Phi)(\rho)$ with
$p\in[1,+\infty]$ provided
$\alpha(\rho)=\operatorname{Tr}H'\Phi(\rho)$. Note that if
$(H\circ\Phi)(\rho_{0})=+\infty$ then \eqref{eq18} may not be valid
(see~\cite{25}, Proposition 7).

By Corollary  \ref{c1} the above coincidence of the convex hulls and
continuity of the Renyi entropy for $p>1$ imply continuity of the
function $\operatorname{co}(R_{p}\circ\Phi)$ for $p>1$ on the convex
subset $\{\rho\in\mathfrak{S}(\mathcal{H})\mid
(H\circ\Phi)(\rho)<+\infty\}$.

If the output von Neumann entropy $H\circ\Phi$ is continuous on a
particular set $\mathcal{A}\subseteq\mathfrak{S}(\mathcal{H})$
then by Theorem 1 in \cite{25} its convex closure
$\overline{\operatorname{co}}(H\circ\Phi)$ is also continuous and
coincides with the convex hull $\operatorname{co}(H\circ\Phi)$ on
this set. If the set $\mathcal{A}$ is compact then the above
assertion on continuity of the function
$\operatorname{co}(R_{p}\circ\Phi)$ and Dini's lemma imply uniform
convergence of the continuous function
$\operatorname{co}(R_{p}\circ\Phi)|_{\mathcal{A}}=\overline{\operatorname{co}}(R_{p}\circ\Phi)|_{\mathcal{A}}$
to the continuous function
$\operatorname{co}(H\circ\Phi)|_{\mathcal{A}}=\overline{\operatorname{co}}(H\circ\Phi)|_{\mathcal{A}}$
as $p\rightarrow1+0$. This shows, in particular, that the Holevo
capacity\footnote{This value is closely related to the classical
capacity of a quantum channel (see \cite{6}).} of the
$\mathcal{A}$-constrained channel $\Phi$ (see~\cite{4}) can be
determined by the expression

$$
\overline{C}(\Phi,\mathcal{A})=\lim_{p\rightarrow 1+0}
\sup_{\rho\in\mathcal{A}}\bigl((R_{p}\circ\Phi)(\rho)-
\operatorname{co}(R_{p}\circ\Phi)(\rho)\bigr).
$$
This expression can be used for approximation of the Holevo capacity
(since the Renyi entropy for $p>1$ is more "computable" than the von
Neumann entropy) and in analysis of continuity of the Holevo
capacity as a function of a channel (since the Renyi entropy is
continuous for $p>1$).
\end{example}

\section{Entanglement monotones}
\label{s4}

\subsection{The basic properties}
\label{s4.1} Entanglement is an essential feature of quantum
systems, which can be considered as a special quantum correlation
having no classical analogue. It is this property that provides a
base for construction of different quantum algorithms and
cryptographic protocols (see~\cite{6}). One of the basic
tasks of the theory of entanglement consists in finding appropriate
quantitative characteristics of entanglement of a state in composite
system and in studying their properties (see~\cite{3},~\cite{27} and
references therein). Entanglement monotones form an important class
of such characteristics \cite{2}. In this section we consider
infinite dimensional generalization of the "convex roof
construction" of entanglement monotones and investigate its
properties. This generalization is based on the results presented in
the previous sections.

Let $\mathcal{H}$ and $\mathcal{K}$~be separable Hilbert spaces. A
state $\omega\in\mathfrak{S}(\mathcal{H}\otimes\mathcal{K})$ is
called \textit{separable} or \textit{nonentangled} if it belongs to
the convex closure of the set of all product pure states in
$\mathfrak{S}(\mathcal{H}\otimes\mathcal{K})$, otherwise it is
called \textit{entangled}.

A key role in the entanglement theory is played by the notion of
\textit{$\mathrm{LOCC}$-operation} in a composite quantum system defined as a composition of Local Operations
on each of the subsystems and Classical Communications between these
subsystems \cite{3}, \cite{27}. Action of a \textit{selective }
$\mathrm{LOCC}$-operation on any state of a composite system results
in a particular \textit{ensemble}~-- collection of states of this system with the
corresponding probability distribution (in general~-- probability
measure on the set of states of this system). A typical example
of a selective $\mathrm{LOCC}$-operation is a quantum measurement on
one of the subsystems, which "transforms" an arbitrary a priory state to
the set of posterior states, corresponding to the outcomes of the
measurement, and the probability distribution of these outcomes
\cite{9}, Ch.\,2. Averaging of the output ensemble of a selective
$\mathrm{LOCC}$-operation gives  the corresponding
\textit{nonselective} $\mathrm{LOCC}$-operation. Thus action of a
nonselective $\mathrm{LOCC}$-operation on any state of a composite
system results in a particular \textit{state} of this system. In the above
example this averaging corresponds to a quantum measurement in which
the result of the measurement is ignored (but a measured state may be
changed).

An \textit{entanglement monotone} is an arbitrary nonnegative
function $E$ on the set
$\mathfrak{S}(\mathcal{H}\otimes\mathcal{K})$ having the following
two properties (see~\cite{2}, \cite{3}).

EM-1) $\{E(\omega)=0\}\Leftrightarrow\{the\;state\;\omega\; is\;
separable\}$.

EM-2a) \textit{Monotonicity of the function $E$ under nonselective
$\mathrm{LOCC}$-operations.} This means that
\begin{equation}
\label{eq19} E(\omega)\ge E\biggl(\sum_{i}\pi_{i}\omega_{i}\biggr)
\end{equation}
for any state
$\omega\in\mathfrak{S}(\mathcal{H}\otimes\mathcal{K})$ and any
$\mathrm{LOCC}$-operation mapping the state $\omega$ to the finite or countable ensemble  $\{\pi_{i},\omega_{i}\}$.

This requirement is often strengthened by the following one.

EM-2b) \textit{Monotonicity of the function $E$ under selective
$\mathrm{LOCC}$-operations.} This means that
\begin{equation}
\label{eq20}
E(\omega)\ge\sum_{i}\pi_{i}E(\omega_{i})
\end{equation}
for any state
$\omega\in\mathfrak{S}(\mathcal{H}\otimes\mathcal{K})$ and any
$\mathrm{LOCC}$-operation mapping the state $\omega$ to the finite or countable ensemble  $\{\pi_{i},\omega_{i}\}$.

In infinite dimensions the last requirement is naturally generalized
to the following one.

EM-2c) \textit{Monotonicity of the function $E$ under generalized
selective $\mathrm{LOCC}$- operations.} This means that for arbitrary
state $\omega\in\mathfrak{S}(\mathcal{H}\otimes\mathcal{K})$ and
local instrument%
\footnote{An \textit{instrument} in the set of states
$\mathfrak{S}(\mathcal{H})$ with a measurable space of outcomes
$\mathcal{X}$ is a set-function $\mathfrak{M}$ defined on the
$\sigma$-algebra $\mathfrak{B}(\mathcal{X})$ satisfying the
following conditions (see \cite{9}, Ch.\,4): $\mathfrak{M}(B)$~is a
linear completely positive trace-nonincreasing transformation of the
space  $\mathfrak{T}(\mathcal{H})$ for any
$B\in\mathfrak{B}(\mathcal{X})$; $\mathfrak{M}(\mathcal{X})$~is
a trace-preserving transformation; if
$\{B_{j}\}\subset\mathfrak{B}(\mathcal{X})$~is a finite or countable
disjoint decomposition of the set $B\in\mathfrak{B}(\mathcal{X})$
then
$\mathfrak{M}(B)[T]=\sum_{j}\mathfrak{M}(B_{j})[T],\,T\in\mathfrak{T}(\mathcal{H})$,
where the series converges in the norm of the space
$\mathfrak{T}(\mathcal{H})$.} $\mathfrak{M}$ with the set of outcomes
$\mathcal{X}$ the function $x\mapsto E(\sigma(x|\,\omega))$ is
$\mu_{\omega}$-measurable on the set $\mathcal{X}$ and
\begin{equation} \label{eq21}
E(\omega)\ge\int_{\mathcal{X}}E(\sigma(x|\,\omega))\mu_{\omega}(dx),
\end{equation}
where
$\mu_{\omega}(\cdot)=\operatorname{Tr}\,\mathfrak{M}(\cdot)[\omega]$
and $\{\sigma(x|\,\omega)\}_{x\in\mathcal{X}}$ are respectively the
probability measure on the set $\mathcal{X}$ describing the results
of the measurement and the family of posteriori states corresponding
to the a priori state $\omega$ \enskip\cite{9}, \cite{28}.

\begin{remark}
\label{r5} By definition the function $x\mapsto \sigma(x|\,\omega)$
is $\mu_{\omega}$-measurable with respect to the minimal
$\sigma$-algebra on $\mathfrak{S}(\mathcal{H}\otimes\mathcal{K})$
for which the all linear functionals
$\omega\mapsto\operatorname{Tr}A\omega$, $A\in\mathfrak{B}(\mathcal{H}\otimes\mathcal{K}),$ are measurable. By
Corollary\,1 in \cite{29} this $\sigma$-algebra coincides with the
Borel $\sigma$-algebra on
$\mathfrak{S}(\mathcal{H}\otimes\mathcal{K})$. Thus the function
$x\mapsto \sigma(x|\,\omega)$ is $\mu_{\omega}$-measurable with
respect to the Borel $\sigma$-algebra on the set
$\mathfrak{S}(\mathcal{H}\otimes\mathcal{K})$ and therefore the function
$x\mapsto E(\sigma(x|\,\omega))$ is $\mu_{\omega}$-measurable for
any Borel function $\omega\mapsto E(\omega)$.
\end{remark}

According to \cite{3} an entanglement monotone $E$ is called \textit{entanglement measure}
if $E(\omega)=H(\operatorname{Tr}_\mathcal{K}\omega)$  for any pure
state $\omega$ in $\mathfrak{S}(\mathcal{H}\otimes\mathcal{K})$,
where $H$ is the von Neumann entropy.

Sometimes the following requirement is included in the definition of
entanglement monotone (cf.~\cite{27}).

EM-3a) \textit{Convexity of the function $E$ on the set
$\,\mathfrak{S}(\mathcal{H}\otimes\mathcal{K})$}, which means that
$$
E\biggl(\sum_{i}\pi_{i}\omega_{i}\biggr)\le
\sum_{i}\pi_{i}E(\omega_{i})
$$
for any finite ensemble $\{\pi_{i},\omega_{i}\}$ of states in
$\mathfrak{S}(\mathcal{H}\otimes\mathcal{K})$.

This requirement is due to the observation that entanglement can not
be increased by taking convex mixtures (describing classical noise
in preparing of a quantum state).

The following two stronger forms of the convexity requirement are
motivated by necessity to consider countable and continuous
ensembles of states dealing with infinite dimensional quantum
systems  (cf.~\cite{4}).

EM-3b) \textit{$\sigma$-convexity of the function $E$ on the set
$\,\mathfrak{S}(\mathcal{H}\otimes\mathcal{K})$}, which means that
$$
E\biggl(\sum_{i}\pi_{i}\omega_{i}\biggr)\le
\sum_{i}\pi_{i}E(\omega_{i})
$$
for any countable ensemble $\{\pi_{i},\omega_{i}\}$ of states in
$\mathfrak{S}(\mathcal{H}\otimes\mathcal{K})$.

If this requirement holds then $\textrm{EM-2b}) \Rightarrow
\textrm{EM-2a})$.

EM-3c) \textit{$\mu$-convexity of the function $E$ on the set
$\,\mathfrak{S}(\mathcal{H}\otimes\mathcal{K})$}, which means that
the function $E$ is universally measurable and
$$
E\biggl(\int_{\mathfrak{S}(\mathcal{H}\otimes\mathcal{K})}\omega\mu(d\omega)\biggr)\le
\int_{\mathfrak{S}(\mathcal{H}\otimes\mathcal{K})}E(\omega)\mu(d\omega)
$$
for any Borel probability measure $\mu$ on the set
$\mathfrak{S}(\mathcal{H}\otimes\mathcal{K})$, which can be
considered as a generalized (continuous) ensemble of states in
$\mathfrak{S}(\mathcal{H}\otimes\mathcal{K})$.

In \S\,\ref{s2} it is shown that these convexity properties are not
equivalent in general.

\goodbreak

EM-4) \textit{Subadditivity  of the function $E$}, which means that
\begin{equation}
\label{eq22}
E(\omega_{1}\otimes\omega_{2})\le E(\omega_{1})+E(\omega_{2})
\end{equation}
for any states
$\omega_{1}\in\mathfrak{S}(\mathcal{H}_{1}\otimes\mathcal{K}_{1})$
and
$\omega_{2}\in\mathfrak{S}(\mathcal{H}_{2}\otimes\mathcal{K}_{2})$.

This property guarantees existence of the regularization
$$
E^{*}(\omega)=\lim_{n\rightarrow+\infty}\frac{E(\omega^{\otimes
n})}{n}, \qquad
\omega\in\mathfrak{S}(\mathcal{H}\otimes\mathcal{K}).
$$

In the finite dimensional case it is natural to require continuity
of an entanglement monotone $E$ on the set
$\mathfrak{S}(\mathcal{H}\otimes\mathcal{K})$. In infinite
dimensions this requirement is very restrictive. Moreover,
discontinuity of the von Neumann entropy implies discontinuity of
any entanglement measure on the set
$\mathfrak{S}(\mathcal{H}\otimes\mathcal{K})$ in this case.
Nevertheless some weaker continuity requirements may be considered.

EM-5a) \textit{Lower semicontinuity of the function $E$ on the set
$\mathfrak{S}(\mathcal{H}\otimes\mathcal{K})$.} This means that
$$
\liminf_{n\rightarrow+\infty}E(\omega_{n})\ge E(\omega_{0})
$$
for any sequence $\{\omega_{n}\}$ of states in
$\mathfrak{S}(\mathcal{H}\otimes\mathcal{K})$ converging to a
state $\omega_{0}$ or, equivalently, that the set of states defined
by the inequality $E(\omega)\le c$ is closed for any $c>0$. This
requirement is motivated by the natural physical observation that
entanglement can not be increased by an approximation procedure. It
is essential that lower semicontinuity of the function $E$ guarantees
that this function is Borel and that requirements EM-3a -- EM-3c are
equivalent for this function (by Proposition~\ref{pA-2} in the
Appendix).

From the physical point of view it is natural to require that
entanglement monotones must be continuous on the set of states produced in
a physical experiment. This leads to the following requirement.

EM-5b) \textit{Continuity of the function $E$ on subsets of
$\,\mathfrak{S}(\mathcal{H}\otimes\mathcal{K})$ with bounded mean
energy.} Let $H_{\mathcal{H}}$ and $H_{\mathcal{K}}$ be the
Hamiltonians of the quantum systems associated with the spaces
$\mathcal{H}$ and $\mathcal{K}$ correspondingly \cite{9}, \S\,1.2.
Then the Hamiltonian of the composite system has the form
$H_{\mathcal{H}}\otimes I_{\mathcal{K}}+I_{\mathcal{H}}\otimes
H_{\mathcal{K}}$ and hence the set of states of the composite system
with the mean energy not exceeding $h$ is defined by the
inequality$$ \operatorname{Tr}(H_{\mathcal{H}}\otimes
I_{\mathcal{K}}+I_{\mathcal{H}}\otimes H_{\mathcal{K}})\omega\le h.
$$

Requirement  EM-5b) means continuity of the restrictions of the
function $E$ to the subsets of
$\mathfrak{S}(\mathcal{H}\otimes\mathcal{K})$ defined by the above
inequality for all $h>0$.

The strongest continuity requirement is the following one.

EM-5c) \textit{Continuity of the function $E$ on the set
$\mathfrak{S}(\mathcal{H}\otimes\mathcal{K})$.}

Despite infinite dimensionality there exists a nontrivial class of
entanglement monotones for which this requirement holds (see Example
\ref{e5} in the next subsection.)

\subsection{The generalized convex roof constructions}

\label{s4.2} In the finite dimensional case a general method of
producing of entanglement monotones is the "convex roof
construction" \cite{3}, \cite{27}, \cite{30}. By this construction
for a given concave continuous nonnegative function $f$ on the set
$\mathfrak{S}(\mathcal{H})$ such that
\begin{equation}
\label{eq23}
f^{-1}(0)=\operatorname{extr}\mathfrak{S}(\mathcal{H}),
\qquad
f(\rho)=f(U\rho U^{*})
\end{equation}
for any state $\rho $ in $\mathfrak{S}(\mathcal{H})$ and
any unitary $U$ in $\mathcal{H}$, the corresponding
entanglement monotone $E^{f}$ is defined as the convex roof
$(f\circ\Theta|_{\operatorname{extr}\mathfrak{S}(\mathcal{H}\otimes\mathcal{K})})_{*}$
of the restriction of the function $f\circ\Theta$ to the set
$\operatorname{extr}\mathfrak{S}(\mathcal{H}\otimes\mathcal{K})$,
where $\Theta\colon
\omega\mapsto\operatorname{Tr}_{\mathcal{K}}\omega$ is a partial
trace. By using the von Neumann entropy in the role of function $f$
in the above construction we obtain the Entanglement of Formation
$E_{F}$ -- one of the most important entanglement measures \cite{7}.

In the infinite dimensional case there exist two possible
generalizations of the above construction: the $\sigma$-convex roof
$(f\circ\Theta|_{\operatorname{extr}\mathfrak{S}(\mathcal{H}\otimes\mathcal{K})})_{*}^{\sigma}$
and the $\mu$-convex roof
$(f\circ\Theta|_{\operatorname{extr}\mathfrak{S}(\mathcal{H}\otimes\mathcal{K})})_{*}^{\mu}$
of the function
$f\circ\Theta|_{\operatorname{extr}\mathfrak{S}(\mathcal{H}\otimes\mathcal{K})}$.
To simplify notations in what follows we will omit the symbol of
restriction and will denote the above functions
$(f\circ\Theta)_{*}^{\sigma}$ and $(f\circ\Theta)_{*}^{\mu}$
correspondingly.

The results of the previous sections make it possible to prove the
following assertions concerning the main properties of these
generalized convex roof constructions.

\begin{theorem}
\label{t2} Let $f$ be a nonnegative concave function on the set
$\mathfrak{S}(\mathcal{H})$ satisfying condition \eqref{eq23}.

{\rm A-1)} If the function $f$ is upper semicontinuous then
$$
(f\circ\Theta)_{*}^{\sigma}=(f\circ\Theta)_{*}^{\mu}=
\mu\textup{-}\!\operatorname{co}(f\circ\Theta)=\sigma\textup{-}\!\operatorname{co}(f\circ\Theta)=\operatorname{co}(f\circ\Theta),
$$
the function $(f\circ\Theta)_{*}^{\mu}=(f\circ\Theta)_{*}^{\sigma}$
is upper semicontinuous and satisfies requirements {\rm EM-1)}, {\rm
EM-2c)} and {\rm EM-3c)}.

{\rm A-2)} If the function $f$ is lower semicontinuous then the
function $(f\circ\Theta)_{*}^{\sigma}$ satisfies requirements%
\footnote{The example in Remark \ref{r6} below shows that the
function $(f\circ\Theta)_{*}^{\sigma}$ may not satisfy requirements
EM-1), EM-3c) and EM-5a) even for bounded lower semicontinuous
function $f$.} {\rm EM-2b)} and {\rm EM-3b)}, while the function
$(f\circ\Theta)_{*}^{\mu}$ coincides with the function
$\overline{\operatorname{co}}(f\circ\Theta)$ and satisfies
requirements {\rm EM-1)}, {\rm EM-2c)}, {\rm EM-3c)} and {\rm
EM-5a)}.

{\rm B)} If the function $f$ is subadditive\footnote{This means that
$f(\rho_{1}\otimes\rho_{2})\le f(\rho_{1})+f(\rho_{2})$ for any
states $\rho_{1}\in \mathfrak{S}(\mathcal{H}_{1})$ and $\rho_{2}\in
\mathfrak{S}(\mathcal{H}_{2})$, where $\mathcal{H}_{1}$ and
$\mathcal{H}_{2}$~are separable Hilbert spaces (we implicitly use
the isomorphism of all such spaces).}, then the functions
$(f\circ\Theta)_{*}^{\sigma}$ and $(f\circ\Theta)_{*}^{\mu}$ satisfy
requirement  {\rm EM-4)}.

{\rm C)} Let $H_{\mathcal{H}}$ be a positive operator in the space
$\mathcal{H}$. If the function $f$ is lower semicontinuous and for each $h>0$ it has
finite continuous restriction to the subset
$\mathcal{K}_{H_{\mathcal{H}},h}=
\{\rho\in\mathfrak{S}(\mathcal{H})\mid
\operatorname{Tr}H_{\mathcal{H}}\rho\le h\}$  then
$$
(f\circ\Theta)_{*}^{\mu}(\omega)=(f\circ\Theta)_{*}^{\sigma}(\omega)=
\overline{\operatorname{co}}(f\circ\Theta)(\omega)=\operatorname{co}(f\circ\Theta)(\omega)
\qquad
\forall\, \omega\in\bigcup_{h>0}\mathcal{K}_{H_{\mathcal{H}}\otimes I_{\mathcal{K}},h},
$$
where $\mathcal{K}_{H_{\mathcal{H}}\otimes I_{\mathcal{K}},h}
=\{\omega\in \mathfrak{S}(\mathcal{H}\otimes\mathcal{K})\mid
\operatorname{Tr}(H_{\mathcal{H}}\otimes I_{\mathcal{K}})\,\omega\le
h\}$, and the common restriction of these functions to the set
$\mathcal{K}_{H_{\mathcal{H}}\otimes I_{\mathcal{K}},h}$ is
continuous for each $h>0$. In particular, if $H_{\mathcal{H}}$ is
the Hamiltonian of the quantum system associated with the space
$\mathcal{H}$ then the functions $(f\circ\Theta)_{*}^{\mu}$ and
$(f\circ\Theta)_{*}^{\sigma}$ satisfy requirement {\rm EM-5b)}.

{\rm D)} If the function $f$ is continuous on the set
$\mathfrak{S}(\mathcal{H})$ then
$$
(f\circ\Theta)_{*}^{\mu}=(f\circ\Theta)_{*}^{\sigma}=
\overline{\operatorname{co}}(f\circ\Theta)=
\mu\textup{-}\!\operatorname{co}(f\circ\Theta)=\sigma\textup{-}\!\operatorname{co}(f\circ\Theta)=\operatorname{co}(f\circ\Theta)
$$
and the function
$(f\circ\Theta)_{*}^{\sigma}=(f\circ\Theta)_{*}^{\mu}$ satisfies
requirement {\rm EM-5c)}.
\end{theorem}

\begin{proof}
A) By Lemma \ref{l2} upper semicontinuity and concavity of the
function $f$ guarantees its boundedness while Proposition \ref{p5}
implies
$$
(f\circ\Theta)_{*}^{\mu}=(f\circ\Theta)_{*}^{\sigma}=\mu\textup{-}\!\operatorname{co}(f\circ\Theta)
=\sigma\textup{-}\!\operatorname{co}(f\circ\Theta)=\operatorname{co}(f\circ\Theta)
$$
and upper semicontinuity of this function. Proposition ~\ref{pA-2} in
the Appendix provides validity of requirement EM-3c) for the
function
$(f\circ\nobreak\Theta)_{*}^{\mu}=(f\circ\Theta)_{*}^{\sigma}$ in
this case.

By Proposition \ref{p3} lower semicontinuity of the function $f$
implies lower semicontinuity of the function
$(f\circ\Theta)_{*}^{\mu}$ (validity of requirement EM-5a). Hence Proposition~\ref{pA-2} in the Appendix
provides validity of requirement EM-3c) for the function
$(f\circ\Theta)_{*}^{\mu}$ in this case.

Validity of requirement EM-3b) for the function
$(f\circ\Theta)_{*}^{\sigma}$ follows from its definition.

By repeating the arguments used in the proof of
$\mathrm{LOCC}$-monotonicity of the convex roof of the function
$f\circ\Theta$ in the finite dimensional case (see~\cite{3},
\cite{7}) and by using discrete Jensen's inequality
(Proposition~\ref{pA-1} in the Appendix) validity of requirement
EM-2b) for the function $(f\circ\Theta)_{*}^{\sigma}$ can be proved.

Consider requirement EM-2c). Let $\mathfrak{M}$ be an arbitrary
instrument acting in the subsystem associated with the space
$\mathcal{K}$. If the function $f$ is lower (correspondingly, upper)
semicontinuous then the function $(f\circ\Theta)_{*}^{\mu}$ is lower
(correspondingly, upper) semicontinuous and hence it is Borel. By
Remark \ref{r5} this guarantees $\mu_{\omega}$-measurability of the
function $x\mapsto E(\sigma(x|\,\omega))$ for any state
$\omega\in\mathfrak{S}(\mathcal{H}\otimes\mathcal{K})$.

Let $\omega$ be a pure state. By locality of the instrument
$\mathfrak{M}$ we have
$$
\Theta(\omega)=\int_{\mathcal{X}}
\Theta(\sigma(x|\,\omega))\mu_{\omega}(dx).
$$
Since the function  $f$ is nonnegative concave and either lower or upper
semicontinuous, Proposition~\ref{pA-2} in the Appendix
implies
$$
f\circ\Theta(\omega)\ge\int_{\mathcal{X}}
f\circ\Theta(\sigma(x|\,\omega))\mu_{\omega}(dx)\ge
\int_{\mathcal{X}}
(f\circ\Theta)_{*}^{\mu}(\sigma(x|\,\omega))\mu_{\omega}(dx),
$$
where the last inequality follows from Proposition \ref{p5}.

Let $\omega$ be a  mixed state. Prove first that
\begin{equation}
\label{eq24}
(f\circ\Theta)_{*}^{\sigma}(\omega)\ge \int_{\mathcal{X}}
(f\circ\Theta)_{*}^{\mu}(\sigma(x|\,\omega))\mu_{\omega}(dx).
\end{equation}
For given $\varepsilon>0$ let $\{\pi_{i},\omega_{i}\}$ be such
ensemble in
$\widehat{\mathcal{P}}_{\{\omega\}}(\mathfrak{S}(\mathcal{H}\otimes\mathcal{K}))$
that
$$
(f\circ\Theta)_{*}^{\sigma}(\omega)>\sum_{i}\pi_{i}f\circ\Theta(\omega_{i})-\varepsilon.
$$
By the above observation concerning a pure state $\omega$ we
have
\begin{equation} \label{eq25}
(f\circ\Theta)_{*}^{\sigma}(\omega)>
\sum_{i}\pi_{i}\int_{\mathcal{X}}
(f\circ\Theta)_{*}^{\mu}(\sigma(x|\,\omega_{i}))\mu_{\omega_{i}}(dx)-\varepsilon.
\end{equation}
By the Radon-Nicodym theorem the decomposition
$$
\mu_{\omega}(\cdot)=\operatorname{Tr}\,\mathfrak{M}(\cdot)[\omega]=\sum_{i}\pi_{i}\operatorname{Tr}\,\mathfrak{M}(\cdot)[\omega_{i}]
=\sum_{i}\pi_{i}\mu_{\omega_{i}}(\cdot)
$$
implies existence of a family $\{p_{i}\}$ of $\mu_{\omega}$-measurable
functions on $\mathcal{X}$ such that
$$
\pi_{i}\mu_{\omega_{i}}(\mathcal{X}_{0})=\int_{\mathcal{X}_{0}}p_{i}(x)\mu_{\omega}(dx)\qquad
$$
for any  $\mu_{\omega}$--measurable subset
$\mathcal{X}_{0}\subseteq\mathcal{X}$ and $\sum_{i}p_{i}(x)=1$ for
$\mu_{\omega}$-almost all $x$ in $\mathcal{X}$. Since
$$
\int_{\mathcal{X}_{0}}\sigma(x|\,\omega)\mu_{\omega}(dx)=
\sum_{i}\pi_{i}\int_{\mathcal{X}_{0}}\sigma(x|\,\omega_{i})\mu_{\omega_{i}}(dx)=
\sum_{i}\int_{\mathcal{X}_{0}}\sigma(x|\,\omega_{i})p_{i}(x)\mu_{\omega}(dx)
$$
for any  $\mu_{\omega}$-measurable subset
$\mathcal{X}_{0}\subseteq\mathcal{X}$ we have
$$
\sum_{i}p_{i}(x)\sigma(x|\,\omega_{i})=\sigma(x|\,\omega)
$$
for $\mu_{\omega}$-almost all $x$ in $\mathcal{X}$.

Note that the function $(f\circ\Theta)_{*}^{\mu}$ is $\sigma$-convex
in the both cases. Indeed, if $f$ is  an upper semicontinuous
function this follows from its coincidence with the function
$(f\circ\Theta)_{*}^{\sigma}$, if $f$ is  a lower semicontinuous
function then the convex function $(f\circ\Theta)_{*}^{\mu}$ is
lower semicontinuous and hence $\mu$-convex (by
Proposition~\ref{pA-2} in the Appendix).

By using \eqref{eq25}  and $\sigma$-convexity of the function
$(f\circ\Theta)_{*}^{\mu}$ we obtain
\begin{align*}
(f\circ\Theta)_{*}^{\sigma}(\omega)
&>\int_{\mathcal{X}}\sum_{i}p_{i}(x)(f\circ\Theta)_{*}^{\mu}(\sigma(x|\,\omega_{i}))
\mu_{\omega}(dx)-\varepsilon
\\
&\ge\int_{\mathcal{X}}(f\circ\Theta)_{*}^{\mu}(\sigma(x|\,\omega))\mu_{\omega}(dx)-\varepsilon,
\end{align*}
which implies \eqref{eq24} since $\varepsilon$ is arbitrary.

If $f$ is an upper semicontinuous function then
$(f\circ\Theta)_{*}^{\sigma}=(f\circ\Theta)_{*}^{\mu}$ and
\eqref{eq24} means \eqref{eq21} for the function
$E=(f\circ\Theta)_{*}^{\sigma}=(f\circ\Theta)_{*}^{\mu}$.

If $f$ is a lower semicontinuous function then for an arbitrary
state $\omega\in\mathfrak{S}(\mathcal{H}\otimes\mathcal{K})$ Lemma
\ref{l1} and Proposition \ref{p5} imply existence of a sequence
$\{\omega_{n}\}\subset\mathfrak{S}(\mathcal{H}\otimes\mathcal{K})$
converging to the state $\omega$ such that
$$
\lim_{n\rightarrow+\infty}(f\circ\Theta)_{*}^{\sigma}(\omega_{n})=(f\circ\Theta)_{*}^{\mu}(\omega).
$$
Inequality \eqref{eq21} for the function
$E=(f\circ\Theta)_{*}^{\mu}$ can be proved by applying inequality
\eqref{eq24} for each state in the sequence $\{\omega_{n}\}$ and
passing to the limit $n\rightarrow+\infty$  by means of Lemma
~\ref{lA-1} in the Appendix and due to lower semicontinuity of the
function $(f\circ\Theta)_{*}^{\mu}$.

Consider requirement EM-1). Note that a state $\omega$ is separable
if and only if there exists a measure $\mu$ in
$\widehat{\mathcal{P}}_{\{\omega\}}(\mathfrak{S}(\mathcal{H}\otimes\mathcal{K}))$
supported by pure product states \cite{14}.

Let $f$ be a lower semicontinuous function. By Proposition \ref{p3}
for an arbitrary state $\omega$ in
$\mathfrak{S}(\mathcal{H}\otimes\mathcal{K})$ there exists a measure
$\mu_{\omega}$ in
$\widehat{\mathcal{P}}_{\{\omega\}}(\mathfrak{S}(\mathcal{H}\otimes\mathcal{K}))$
such that $(f\circ\Theta)_{*}^{\mu}(\omega)=\int
f\circ\Theta(\sigma)\mu_{\omega}(d\sigma)$. Hence validity of
requirement EM-1) for the function $(f\circ\Theta)_{*}^{\mu}$
follows from the above characterization of the set of separable
states.

Let $f$ be an upper semicontinuous function. Then the function $(f\circ\Theta)_{*}^{\sigma}=(f\circ\Theta)_{*}^{\mu}$ equals
to zero on the set of separable states by the above characterization
of this set.

Suppose this function equals to zero at some entangled state
$\omega_{0}$. Then there exists a local operation $\Lambda$ such that
the state $\Lambda(\omega_{0})$ is entangled and has reduced states
of finite rank. By $\mathrm{LOCC}$-monotonicity of the function
$(f\circ\Theta)_{*}^{\sigma}=(f\circ\Theta)_{*}^{\mu}$ proved before
this function equals to zero at the entangled state
$\Lambda(\omega_{0})$.

Let $\mathcal{H}_{0}$ be the finite dimensional support of the state
$\operatorname{Tr}_{\mathcal{K}}\Lambda(\omega_{0})$. Then the upper
semicontinuous concave function $f$ satisfying condition
\eqref{eq23}  has continuous restriction to the set
$\mathfrak{S}(\mathcal{H}_{0})$. Indeed, continuity of this
restriction at any pure state in $\mathfrak{S}(\mathcal{H}_{0})$
follows from upper semicontinuity of the nonnegative function $f$
and condition \eqref{eq23}, while continuity of this restriction at
any mixed state in $\mathfrak{S}(\mathcal{H}_{0})$ can be easily
derived from the well known fact that any concave bounded function
is continuous at any internal point of a convex subset of a Banach
space \cite{21}, Proposition 3.2.3. Since
$$
(f\circ\Theta|_{\mathfrak{S}(\mathcal{H}_{0}\otimes\mathcal{K})})_{*}^{\mu}=
(f\circ\Theta)_{*}^{\mu}|_{\mathfrak{S}(\mathcal{H}_{0}\otimes\mathcal{K})},
$$
we can apply the previous observation concerning lower
semicontinuous function $f$ to show that equality
$(f\circ\Theta)_{*}^{\mu}(\Lambda(\omega_{0}))=0$ implies
separability of the state $\Lambda(\omega_{0})$, contradicting to
the above assumption.

B) If the function $f$ is subadditive then the function
$f\circ\Theta$ is subadditive as well. Let
$\mu_{i}\in\widehat{\mathcal{P}}_{\{\omega_{i}\}}(\mathfrak{S}(\mathcal{L}_{i}))$,
where $\mathcal{L}_{i}=\mathcal{H}_{i}\otimes\mathcal{K}_{i}$,
$i=1,2$,~be arbitrary measures. The set of product states
in~$\operatorname{extr}\mathfrak{S}(\mathcal{L}_{1}\otimes\mathcal{L}_{2})$
can be considered as the Cartesian product of the sets
$\operatorname{extr}\mathfrak{S}(\mathcal{L}_{1})$ and
$\operatorname{extr}\mathfrak{S}(\mathcal{L}_{2})$. Hence on this
set one can define the Cartesian product of the measures $\mu_{1}$
and $\mu_{2}$, denoted by $\mu_{1}\otimes\mu_{2}$, which can be
considered as a measure in
$\widehat{\mathcal{P}}_{\{\omega_{1}\otimes\omega_{2}\}}
(\mathfrak{S}(\mathcal{L}_{1}\otimes\mathcal{L}_{2}))$ supported by
the set of product states. By using this construction it is easy to
prove subadditivity of the function $(f\circ\Theta)_{*}^{\mu}$. By
the same argumentation with atomic measures $\mu_{1}$ and $\mu_{2}$ one
can prove\footnote{In this case the measure $\mu_{1}\otimes\mu_{2}$
corresponds to the tensor product of countable ensembles of pure
states corresponding to the measures $\mu_{1}$ and $\mu_{2}$.}
subadditivity of the function $(f\circ\Theta)_{*}^{\sigma}$.

C) If the function $f$ is lower semicontinuous and satisfies the
additional conditions in assertion C of the theorem, then the
function $f\circ\Theta$ satisfies the conditions of Corollary
\ref{c5} with the affine function
$\alpha(\omega)=\operatorname{Tr}(H_{\mathcal{H}}\otimes
I_{\mathcal{K}})\omega$.

D) Assertion D follows from Proposition \ref{p5}.
\end{proof}

\begin{remark}
\label{r6} The function $(f\circ\Theta)_{*}^{\sigma}$ may not
satisfy the basic requirement EM-1) even for bounded lower
semicontinuous function $f$ (see assertion ~A-2) of
Theorem~\ref{t2}). Indeed, let $f$ be the indicator function of the
set of all mixed states in $\mathfrak{S}(\mathcal{H})$ and
$\omega_{0}$ be a separable state such that any measure in
$\widehat{\mathcal{P}}_{\{\omega_{0}\}}(\mathfrak{S}(\mathcal{H}\otimes\mathcal{K}))$
has no atoms within the set of separable states \cite{14}. Then it is
easy to see that $(f\circ\Theta)_{*}^{\sigma}(\omega_{0})=1$ (while
$(f\circ\Theta)_{*}^{\mu}(\omega_{0})=0$).
\end{remark}

The  function $(f\circ\Theta)_{*}^{\sigma}$ in Remark~\ref{r6}
does not also satisfy requirements EM-3c) and EM-5a). This is a
general feature of any $\sigma$-convex roof not coinciding with the
corresponding $\mu$-convex roof.

Remark \ref{r6} and Theorem \ref{t2}  show that the function
$(f\circ\Theta)_{*}^{\sigma}$ either coincides with the function
$(f\circ\Theta)_{*}^{\mu}$ (if $f$ is upper semicontinuous) or may
not satisfy the basic requirement EM-1 of entanglement monotones (if
$f$ is lower semicontinuous). Thus the $\mu$-convex roof
construction seems to be more \textit{preferable} candidate on the
role of infinite dimensional generalization of the convex roof
construction of entanglement monotones. Thus we will use the
following notation:
$$
E^{f}=(f\circ\Theta)_{*}^{\mu}
$$
for any function $f$ satisfying the conditions of Theorem
\ref{t2}.

\begin{example}
\label{e5} Generalizing to the infinite dimensional case the
observation in \cite{30} consider the family of functions
$$
f_{\alpha}(\rho)=2(1-\operatorname{Tr}\rho^{\alpha}),
\qquad \alpha>1,
$$
on the set $\mathfrak{S}(\mathcal{H})$ with
$\dim\mathcal{H}=+\infty$. The functions of this family are
nonnegative concave continuous and satisfy conditions \eqref{eq23}.
By Theorem \ref{t2} $E^{f_{\alpha}}$ is an entanglement monotone,
satisfying requirements  EM-1), EM-2c), EM-3c) and EM-5c). In the case
$\alpha=2$ the entanglement monotone $E^{f_{2}}$ can be considered
as the infinite dimensional generalization of the I-tangle
\cite{31}. By Corollary \ref{c4} the function
$(\omega,\alpha)\mapsto E^{f_{\alpha}}(\omega)$ is continuous on the
set $\mathfrak{S}(\mathcal{H}\otimes\mathcal{K})\times[1,+\infty)$.
By Corollary \ref{c3} the least upper bound of the monotonic family
$\{E^{f_{\alpha}}\}_{\alpha>1}$ of continuous entanglement monotones
coincides with the indicator function of the set of entangled
states.
\end{example}

\begin{example}
\label{e6} Let $R_{p}(\rho)=\frac{\log\operatorname{Tr}\rho^{p}}{
1-p}$~be the Renyi entropy of the state
$\rho\in\mathfrak{S}(\mathcal{H})$ of order $p\in[0,1]$ (the case
$p=0$ corresponds to the function $\log\operatorname{rank}(\rho)$,
the case $p=1$ corresponds to the von Neumann entropy), $R_{p}$~--
is a concave lower semicontinuous subadditive function on the set
$\mathfrak{S}(\mathcal{H})$ with the range $[0,+\infty]$, satisfying
condition \eqref{eq23}. By Theorem \ref{t2} the function $E^{R_{p}}$
is an entanglement monotone, satisfying requirements EM-1), EM-2c),
EM-3c), EM-4) and EM-5a). In the case $p=0$ the entanglement
monotone $E^{R_{0}}$ is an infinite dimensional generalization of
the Schmidt measure \cite{27}. In the case $p=1$ the entanglement
monotone $E^{R_{1}}=E^{H}$ is an entanglement measure, which can be
considered as an infinite dimensional generalization of the
Entanglement of Formation  \cite{7} (see the next section). If
$\operatorname{g}(H_{\mathcal{H}})=\inf\{\lambda>0\mid
\operatorname{Tr}\exp(-\lambda H_{\mathcal{H}})<+\infty \}=0$ then
Theorem \ref{t2},\,C) implies that the entanglement measure
$E^{R_{1}}=E^{H}$ satisfies requirement EM-5b), since the von
Neumann entropy $H=R_{1}$ is continuous on the set
$\mathcal{K}_{H_{\mathcal{H}},h}$ (see~\cite{12} or \cite{26}, Proposition 1).
\end{example}

\subsection{Approximation of entanglement monotones}
\label{s4.3} In general entanglement monotones produced by the
$\mu$-convex roof construction are unbounded and discontinuous (only
lower or upper semicontinuous), which may lead to analytical
problems in dealing with these functions. Some of these problems can
be solved by using the following approximation result.

\begin{proposition}
\label{p7} Let $f$ be a concave nonnegative lower semicontinuous
(correspondingly, upper semicontinuous) function on the set
$\,\mathfrak{S}(\mathcal{H})$ satisfying condition \eqref{eq23},
which is represented as a pointwise limit of some increasing (correspondingly,
decreasing) sequence $\{f_{n}\}$ of concave continuous nonnegative
functions on the set $\,\mathfrak{S}(\mathcal{H})$ satisfying
condition \eqref{eq23}. Then the entanglement monotone $E^{f}$ is a
pointwise limit of the increasing (correspondingly, decreasing) sequence
$\{E^{f_{n}}\}_{n}$ of continuous entanglement monotones.

If, in addition, the function $f$ satisfies condition C in Theorem
\ref{t2}, then the sequence $\{E^{f_{n}}\}$ converges to the
entanglement monotone $E^{f}$ uniformly on compact subsets of the
set $\mathcal{K}_{H_{\mathcal{H}}\otimes I_{\mathcal{K}},h}$ for
each $h>0$.
\end{proposition}

\begin{proof}
The first assertion of this proposition follows from Theorem
\ref{t2}, Corollary \ref{c3} and Remark \ref{r2}. The second
assertion follows from the first one and Dini's lemma.
\end{proof}

\section{Entanglement of Formation}
\label{s5}

\subsection{The two definitions}
\label{s5.1} The Entanglement of Formation (EoF) of a state $\omega$
of a finite dimensional composite system is defined in~\cite{7} as
the minimal possible average entanglement over all pure state
\textit{discrete finite} decompositions of $\omega$ (entanglement of a
pure state is defined as the von Neumann entropy of its reduced
state). In our notations this means that
$$
E_{F}=(H\circ\Theta)_{*}=\overline{\operatorname{co}}(H\circ\Theta)=\operatorname{co}(H\circ\Theta).
$$

The possible generalization of this notion is considered
in~\cite{8}, where the Entanglement of Formation of a state $\omega$
of an infinite dimensional composite system is defined as the
minimal possible average entanglement over all pure state
\textit{discrete countable} decompositions of $\omega$,  which means
$E^{d}_{F}=(H\circ\Theta)_{*}^{\sigma}$.

The  generalized convex roof construction considered in Section~\ref{s4.2}
with the von Neumann entropy $H$ in the role of function $f$ leads
to the proposed in~\cite{25} definition of the EoF:
$E^{c}_{F}=E^{H}=(H\circ\Theta)_{*}^{\mu}=\overline{\operatorname{co}}(H\circ\Theta)$,
by which the Entanglement of Formation of a
state $\omega$ of an infinite dimensional composite system is
defined as the minimal possible average entanglement over all pure
state \textit{continuous} decompositions of $\omega$.

An interesting open question is a relation between $E^{d}_{F}$ and
$E^{c}_{F}$. It follows from the definitions that
$$
E^{d}_{F}(\omega)\ge E^{c}_{F}(\omega)
\qquad \forall\,
\omega\in\mathfrak{S}(\mathcal{H}\otimes\mathcal{K}).
$$
In  \cite{25} it is shown that
\begin{equation}
\label{eq26}
E^{d}_{F}(\omega)=E^{c}_{F}(\omega)
\end{equation}
for any state $\omega$ such that either
$H(\operatorname{Tr}_{\mathcal{H}}\omega)<+\infty$ or
$H(\operatorname{Tr}_{\mathcal{K}}\omega)<+\infty$. Equality
\eqref{eq26} obviously holds for all pure states and for all
nonentangled states, but its validity for arbitrary state $\omega$
is not proved (as far as I know). The example in Remark \ref{r6} shows
that this question can not be solved by using only such analytical
properties of the von Neumann entropy as concavity and lower
semicontinuity. Note that the question of coincidence of the
functions $E^{d}_{F}$ and $E^{c}_{F}$ is equivalent to the question
of lower semicontinuity of the function $E^{d}_{F}$, since
$E^{c}_{F}$ is the greatest lower semicontinuous convex function
coinciding with the von Neumann entropy on the set of pure states.

Despite the fact that the definition of the function $E^{d}_{F}$
seems more reasonable from the physical point of view (since it
involves optimization over ensembles of quantum states rather then
measures) the assumption of existence of a state $\omega_{0}$ such
that $E^{d}_{F}(\omega_{0})\neq E^{c}_{F}(\omega_{0})$ leads to the
following "nonphysical" property of the function $E^{d}_{F}$. For
each natural $n$ consider the local measurement $\{M^{n}_{k}\}_{k\in
\mathbb{N}}$, where
$$
M_{1}=\biggl(\sum_{i=1}^{n}|i\rangle\langle i|\biggr)\otimes I_{\mathcal{K}},
\qquad
M_{k}=|n+k-1\rangle\langle n+k-1|\otimes I_{\mathcal{K}},
\quad k>1.
$$
It is clear that the sequence $\{\Phi_{n}\}_{n}$, where
$\Phi_{n}=\{M^{n}_{k}\}_{k\in \mathbb{N}}$, of nonselective local
operations tends to the trivial operation -- the identity
transformation (in the strong operator topology). Since the functions $E^{d}_{F}$ and $E^{c}_{F}$
satisfy requirement EM-2b) and EM-3b), for each $n$ we have
\begin{gather*}
E^{d}_{F}(\omega_{0})
\ge \sum_{k=1}^{+\infty}\pi^{n}_{k}E^{d}_{F}(\omega^{n}_{k})
\ge E^{d}_{F}\biggl(\sum_{k=1}^{+\infty}\pi^{n}_{k}\omega^{n}_{k}\biggr)=E^{d}_{F}(\Phi_{n}(\omega_{0})),
\\
E^{c}_{F}(\omega_{0})\ge
\sum_{k=1}^{+\infty}\pi^{n}_{k}E^{c}_{F}(\omega^{n}_{k}),
\end{gather*}
where $\pi^{n}_{k}=\operatorname{Tr}M^{n}_{k}\omega_{0}M^{n}_{k}$~is
the probability of $k$-th outcome and
$\omega^{n}_{k}=(\pi^{n}_{k})^{-1}M^{n}_{k}\omega_{0}M^{n}_{k}$~is
the posteriori state corresponding to this outcome \cite{9}, Ch.~4.

Since for each $n$ and $k$ the state
$\operatorname{Tr}_{\mathcal{K}}\omega^n_{k}$ has finite rank, the
above-mentioned result in \cite{25} implies
$E^{d}_{F}(\omega^n_{k})=E^{c}_{F}(\omega^n_{k})$. Thus the above two
inequalities show that
$$ E^{d}_{F}(\Phi_{n}(\omega_{0}))
=E^{d}_{F}\biggl(\sum_{k=1}^{+\infty}\pi^{n}_{k}\omega^{n}_{k}\biggr)\le
E^{c}_{F}(\omega_{0})
$$
for all $n$ and hence,
$$
\limsup_{n\rightarrow+\infty}E^{d}_{F}(\Phi_{n}(\omega_{0}))\le
E^{d}_{F}(\omega_{0})-\Delta,
\qquad
\Delta=E^{d}_{F}(\omega_{0})-E^{c}_{F}(\omega_{0})>0,
$$
despite the fact that the sequence $\{\Phi_{n}\}_{n}$ of
nonselective local operations tends to the identity
transformation. In contrast to this lower semicontinuity and
$\mathrm{LOCC}$-monotonicity of the function $E^{c}_{F}$ implies
$$
\lim_{n\rightarrow+\infty}E^{c}_{F}(\Phi_{n}(\omega_{0}))=E^{c}_{F}(\omega_{0})
$$
for any state $\omega_{0}$ and any sequence
$\{\Phi_{n}\}_{n}$ of nonselective $\mathrm{LOCC}$-operations
tending to the identity transformation.

The another advantage of the function $E^{c}_{F}$ consists in its
generalized $\mathrm{LOCC}$-monotonicity (validity of
requirements EM-2c))  following from Theorem \ref{t2},
while the assumption $E^{d}_{F}\neq E^{c}_{F}$ means that the
function $E^{d}_{F}$ is not lower semicontinuous, which is a real
obstacle to prove the analogous property for this function.

\subsection{The approximation of EoF}
\label{s5.2} For given natural $n>1$ consider the function $H_{n}$
on the set $\mathfrak{S}(\mathcal{H})$ defined as follows
$$
H_{n}(\rho)=\sup\sum_{i}\pi_{i}H(\rho_{i}),
$$
where the supremum is over all countable ensembles
$\{\pi_{i},\rho_{i}\}$ of states of rank $\leq n$ such that
$\sum_{i}\pi_{i}\rho_{i}=\rho$. It is easy to see that the function
$H_{n}$ is concave, satisfies condition
\eqref{eq23}, has the range $[0,\log n]$ and coincides with the von
Neumann entropy on the subset of $\mathfrak{S}(\mathcal{H})$
consisting of states of rank $\le n$. By using the strengthened
version of the stability property of the set
$\mathfrak{S}(\mathcal{H})$ in~\cite{32} it is shown  that the
function $H_{n}$ is continuous on the set
$\mathfrak{S}(\mathcal{H})$ and that the increasing sequence
$\{H_{n}\}$ pointwise converges to the von Neumann entropy on this
set.

By Theorem \ref{t2} the function
$E^{n}_{F}=(H_{n}\circ\Theta)_{*}^{\mu}$ is an entanglement monotone
satisfying requirements EM-1), EM-2c), EM-3c), EM-4) and EM-5c). It
is easy to see that the function $E^{n}_{F}$ has the range $[0,\log
n]$ and coincides with the function $E^{c}_{F}$ on the set
$$
\bigl\{\omega\in\mathfrak{S}(\mathcal{H}\otimes\mathcal{K})\mid
\min\{\operatorname{rank}\operatorname{Tr}_{\mathcal{K}}\omega,
\operatorname{rank}\operatorname{Tr}_{\mathcal{H}}\omega\}\le
n\bigr\}.
$$

By Proposition \ref{p7} the sequence $\{E^{n}_{F}\}$ provides
approximation of the function $E^{c}_{F}$ on the set
$\mathfrak{S}(\mathcal{H}\otimes\mathcal{K})$, which is uniform on
each compact set of continuity of the function $E^{c}_{F}$, in
particular, on compact subsets of the set
$\mathcal{K}_{H_{\mathcal{H}}\otimes I_{\mathcal{K}},h}$ for all
 $h>0$, where $H_{\mathcal{H}}$~is a
$\mathfrak{H}$-operator in the space $\mathcal{H}$ such that
$\operatorname{Tr}e^{-\lambda H_{\mathcal{H}}}<+\infty$ for any
$\lambda>0$. Conditions of continuity of the function  $E^{c}_{F}$
are considered in the next subsection.

\subsection{Continuity conditions for EoF}
\label{s5.3} Theorem 1 in \cite{25} implies the following
continuity condition for the function $E^{c}_{F}$, which can be also
formulated as a continuity condition for the function $E^{d}_{F}$,
since this condition implies coincidence of these functions.

\begin{proposition}
\label{p8} The function $E^{c}_{F}$ has continuous restriction to
a set
$\mathcal{A}\subset\mathfrak{S}(\mathcal{H}\otimes\mathcal{K})$ if
either the function $\omega\mapsto
H(\operatorname{Tr}_{\mathcal{H}}\omega)$ or the function
$\omega\mapsto H(\operatorname{Tr}_{\mathcal{K}}\omega)$ has
continuous restriction to the set $\mathcal{A}$.
\end{proposition}
This condition implies the result mentioned in Example \ref{e6}
(validity of requirement EM-5b) as well as the following
observation.

\begin{corollary}
\label{c8} Let $\rho$ be a state in $\mathfrak{S}(\mathcal{H})$. The function $E^{c}_{F}$ has continuous restriction to
the set $\{\omega\mid \operatorname{Tr}_{\mathcal{K}}\omega=\rho\}$ if and only if $H(\rho)<+\infty$.
\end{corollary}

\begin{proof}
It is sufficient to note that if $H(\rho)=+\infty$ then there exists a
pure state $\omega\in\mathfrak{S}(\mathcal{H}\otimes\mathcal{K})$
such that $\operatorname{Tr}_{\mathcal{K}}\omega=\rho$.
\end{proof}

By Corollary \ref{c8} for arbitrary continuous family
$\{\Psi_{t}\}_{t}$ of local operations on the quantum system
associated with the space $\mathcal{K}$ and arbitrary state
$\omega\in\mathfrak{S}(\mathcal{H}\otimes\mathcal{K})$ such that
$\operatorname{Tr}_{\mathcal{K}}\omega<+\infty$ the function
$t\mapsto E^{c}_{F}(\Psi_{t}(\omega))$ is continuous.

For an arbitrary state $\sigma$ let
$\operatorname{d}(\sigma)=\inf\{\lambda\in\mathbb{R}\mid
\operatorname{Tr}\sigma^{\lambda}<+\infty\}$ be the characteristic
of the spectrum of this state. It is clear that
$\operatorname{d}(\sigma)\in [0,1]$. Proposition\,\ref{p8},
Proposition\,2 in \cite{26} and the monotonicity of the
relative entropy imply the following condition of continuity of the
function $E^{c}_{F}$ with respect to the convergence defined by the
relative entropy (which is stronger than the convergence
defined by the trace norm).

\begin{corollary}
\label{c9} Let $\omega_{0}$ be a state in
$\mathfrak{S}(\mathcal{H}\otimes\mathcal{K})$ such that either
$\operatorname{d}(\operatorname{Tr}_{\mathcal{H}}\omega)<\nobreak1$
or $\operatorname{d}(\operatorname{Tr}_{\mathcal{K}}\omega)<1$. If
$\{\omega_{n}\}$ is a sequence such that
$\lim_{n\rightarrow+\infty}H(\omega_{n}\|\omega_{0})=0$ then
$\lim_{n\rightarrow+\infty}E^{c}_{F}(\omega_{n})=E^{c}_{F}(\omega_{0})$.
\end{corollary}

\section{Possible generalizations}
\label{s6}

The definitions of $\sigma$-convexity and  $\mu$-convexity are
naturally generalized to functions defined on an arbitrary convex
closed subset of a locally convex space if any probability measure on
this set has a well defined barycenter. The definitions of
$\sigma$-convex and $\mu$-convex roofs also admit such
generalizations but it is necessary to impose conditions providing
correctness of these constructions.

There exists a class of convex subsets of locally convex spaces
including all metrizable compact sets as well as several noncompact
sets (in particular, the set $\mathfrak{S}(\mathcal{H})$ of quantum
states), to which the main results obtained in
~\S\,\ref{s2},~\ref{s3} can be extended. This class of subsets
called $\mu$-compact in~\cite{13} is studied in detail in \cite{15},
where possibility to extend several results well
known for convex compact sets (in particular, the
Choquet theorem on barycenter decomposition and the
Versterstrem-O'Brien theorem) to $\mu$-compact sets is shown \cite{16}. The last theorem states
equivalence of the stability property of a convex $\mu$-compact set
(which means openness of the convex mixture map) and several other
properties, in particular, openness of the barycenter map and
openness of the restriction of this map to the set of measures
supported by extreme points.

By using results in  \cite{13}, \cite{15} it is easy to show that the
all assertions in \S\,\ref{s2}, \ref{s3} are valid for arbitrary
convex stable $\mu$-compact set $\mathcal{A}$ (instead of
$\mathfrak{S}(\mathcal{H})$) such that
$\mathcal{A}=\sigma\textrm{-}\operatorname{co}(\operatorname{extr}\mathcal{A})$.
Stability of $\mathcal{A}$ is used only in the proofs of Propositions
\ref{p2}, \ref{p4}, Corollaries  \ref{c1}, \ref{c2}, \ref{c4}, the
second part of Proposition \ref{p5}, Theorem \ref{t1} and its
corollaries while in the proofs of Proposition \ref{p3} and
Corollary \ref{c3} it can be replaced by the weaker requirement of
closedness of the set $\operatorname{extr}\mathcal{A}$, which is
necessary for definition of the $\mu$-convex roof. The condition
$\mathcal{A}=\sigma\textrm{-}\operatorname{co}(\operatorname{extr}\mathcal{A})$
is necessary for definition of the $\sigma$-convex roof and is used
in the proofs of all assertions related with this construction.

\section*{Appendix}

{\bf A1. Jensen's inequalities for functions on Banach spaces.} Here
sufficient conditions for validity of Jensen's inequality (in
descrete and integral forms) for convex functions on Banach spaces
taking values in ~$[-\infty,+\infty]$ are presented. As a simple
example showing importance of the conditions in the below
propositions one can consider the affine Borel function on the simplex
of all probability distributions with countable number of outcomes
taking the value $0$ on finite rank distributions and the value
$+\infty$ on infinite rank distributions. Other examples are
considered in~\S\,\ref{s2}.

By using Jensen's inequality for finite convex combinations and
a simple approximation it is easy to prove the following assertion.

\begin{prop}[{\rm (discrete Jensen's inequality)}]
\label{pA-1} Let $f$ be a convex upper bounded function on a closed
convex bounded subset $\mathcal{A}$ of a Banach space. Then for
arbitrary countable set $\{x_{i}\}\subset\mathcal{A}$ with the
corresponding probability distribution $\{\pi_{i}\}$  the following
inequality holds
$$
f\biggl(\sum_{i=1}^{+\infty}\pi_{i}x_{i}\biggr)\le
\sum_{i=1}^{+\infty}\pi_{i}f(x_{i}).
$$
\end{prop}

\begin{prop}[{\rm (integral Jensen's inequality)}]
\label{pA-2} Let $f$ be a convex function on a closed bounded convex
subset $\mathcal{A}$ of a separable Banach space which is either
lower semicontinuous or upper bounded and upper semicontinuous. Then for
arbitrary Borel probability measure $\mu$ on the set $\mathcal{A}$
the following inequality holds
\begin{equation} \label{eq27}
f\biggl(\int_{\mathcal{A}}x\mu(dx)\biggr)\le
\int_{\mathcal{A}}f(x)\mu(dx).
\end{equation}
\end{prop}
(If $\mathcal{A}$~is a subset in $\mathbb{R}^{n}$ then inequality
\eqref{eq27} holds for any Borel function  $f$ taking values in
~$[-\infty,+\infty]$ and any Borel measure $\mu$\enskip \cite{33}.)

\begin{proof} Let $\mu_{0}$ be an arbitrary probability measure
on the set $\mathcal{A}$.

Let $f$ be an upper bounded upper semicontinuous function. Then the
functional $\mu\mapsto\int_{\mathcal{A}}f(x)\mu(dx)$ is upper
semicontinuous on the set $\mathcal{P}(\mathcal{A})$ of Borel
probability measures on $\mathcal{A}$ endowed with the weak convergence
topology \cite{24}, \S\,2. Let $\{\mu_{n}\}$ be a sequence of
measures with finite support and the same barycenter as the measure
$\mu_{0}$ weakly converging to the measure $\mu_{0}$. By convexity
of the function $f$ inequality \eqref{eq27} holds with $\mu=\mu_{n}$
for each $n$. By upper semicontinuity of the functional
$\mu\mapsto\int_{\mathcal{A}}f(x)\mu(dx)$ passing to the limit
$n\rightarrow+\infty$ in this inequality implies inequality
\eqref{eq27} with~$\mu=\mu_{0}$.

Let $f$ be a lower semicontinuous function. By using the arguments
from the proof of Lemma \ref{l2} one can show that the function $f$
is either lower bounded or does not take finite values. It is
sufficient to consider the first case. Suppose that
$\int_{\mathcal{A}}f(x)\mu(dx)<+\infty$. By applying the
construction used in the proof of Lemma \ref{l1} it is possible to
obtain a sequence $\{\mu_{n}\}$ of measures on the set $\mathcal{A}$
with finite support such that
$$
\limsup_{n\rightarrow+\infty}\int_{\mathcal{A}}f(x)\mu_{n}(dx)\le\!
\int_{\mathcal{A}}f(x)\mu_{0}(dx),
\qquad
\lim_{n\rightarrow+\infty}\int_{\mathcal{A}}
x\mu_{n}(dx)=\int_{\mathcal{A}}x\mu_{0}(dx).
$$
By convexity of the function $f$ inequality \eqref{eq27} holds with
$\mu=\mu_{n}$ for each $n$. By lower semicontinuity of the function
$f$ passing to the limit $n\rightarrow+\infty$ implies inequality
\eqref{eq27} with $\mu=\mu_{0}$.
\end{proof}

\begin{corol}
\label{cA-1} Let $f$ be an affine lower semicontinuous function on a
closed bounded convex subset $\mathcal{A}$ of a separable Banach
space. Then for arbitrary Borel probability measure $\mu$ on the set
$\mathcal{A}$ the following equality holds
\begin{equation}
\label{eq28} f\biggl(\int_{\mathcal{A}}x\mu(dx)\biggr)=
\int_{\mathcal{A}}f(x)\mu(dx).
\end{equation}
\end{corol}

{\bf A2. One property of posteriori states.} Let $\mathfrak{M}$ be
an arbitrary instrument on the set $\mathfrak{S}(\mathcal{H})$ with
the set of outcomes $\mathcal{X}$\enskip \cite{9}, Ch.\,4. For a
given arbitrary state $\rho\in\mathfrak{S}(\mathcal{H})$ let
$\mu_{\rho}(\cdot)=\operatorname{Tr}\mathfrak{M}(\cdot)[\rho]$ be
the posteriori measure on the set $\mathcal{X}$ and
$\{\sigma(x|\rho)\}_{x\in \mathcal{X}}$ be the family of
posteriori states corresponding to the a priori state $\rho$\enskip
\cite{9}, \cite{28}.

\begin{lemm}
\label{lA-1} For arbitrary convex lower semicontinuous function $f$
on the set $\mathfrak{S}(\mathcal{H})$ and arbitrary sequence
$\{\rho_{n}\}\subset\mathfrak{S}(\mathcal{H})$ converging to a state
$\rho_{0}$ the following relation holds
$$
\liminf_{n\rightarrow+\infty}\int_{\mathcal{X}}f(\sigma(x|\rho_{n}))\mu_{\rho_{n}}(dx)\ge
\int_{\mathcal{X}}f(\sigma(x|\rho_{0}))\mu_{\rho_{0}}(dx).
$$
\end{lemm}

\begin{proof}
It is sufficient to show that the assumption
\begin{equation}
\label{eq29}
\lim_{n\rightarrow+\infty}\int_{\mathcal{X}}f(\sigma(x|\rho_{n}))\mu_{\rho_{n}}(dx)\le
\int_{\mathcal{X}}f(\sigma(x|\rho_{0}))\mu_{\rho_{0}}(dx)-\Delta,\qquad
\Delta>0,
\end{equation}
leads to a contradiction.

Let $\nu_{0}=\mu_{\rho_{0}}\circ\,\sigma^{-1}(\,\cdot\,|\rho_{0})$
be the image of the measure $\mu_{\rho_{0}}$ under the map
$x\mapsto\sigma(x|\rho_{0})$. It is clear that
$\nu_{0}\in\mathcal{P}$ (see Remark \ref{r5}) and that
$$
\int_{\mathcal{X}}f(\sigma(x|\rho_{0}))\mu_{\rho_{0}}(dx)
=\int_{\mathfrak{S}(\mathcal{H})}f(\rho)\nu_{0}(d\rho).
$$
By separability of the set $\mathfrak{S}(\mathcal{H})$ for given $m$
one can find a family $\{\mathcal{B}_{i}^{m}\}_{i}$ of Borel subsets
of $\mathfrak{S}(\mathcal{H})$ such that
$\nu_{0}(\mathcal{B}_{i}^{m})>0$ for all $i$ and the sequence of
measures
$$ \nu_{m}=\left\{\nu_{0}(\mathcal{B}_{i}^{m}),
\frac{1}{\nu_{0}(\mathcal{B}_{i}^{m})}\int_{\mathcal{B}_{i}^{m}}\rho
\nu_{0}(d\rho)\right\}_{i}
$$
weakly converges to the measure $\nu_{0}$ (see the proof of
Lemma 1 in \cite{4}). Lower semicontinuity of the functional
$\mu\mapsto\int_{\mathfrak{S}(\mathcal{H})}f(\rho)\mu (d\rho)$
implies existence of such $m_{0}$ that
\begin{align}
\nonumber
&\sum_{i}\nu_{0}(\mathcal{B}_{i}^{m_{0}})f
\biggl(\frac{1}{\nu_{0}(\mathcal{B}_{i}^{m_{0}})}\int_{\mathcal{B}_{i}^{m_{0}}}\rho
\nu_{0}(d\rho)\biggr)
\\
&\qquad\qquad\qquad=
\int_{\mathfrak{S}(\mathcal{H})}f(\rho)\nu_{m_{0}}(d\rho)\ge
\int_{\mathfrak{S}(\mathcal{H})}f(\rho)\nu_{0}(d\rho)-\frac{1}{3}\Delta.
\label{eq30}
\end{align}
By using the finite family $\{\mathcal{X}_{i}\}$,
$\mathcal{X}_{i}=\sigma^{-1}(\mathcal{B}_{i}^{m_{0}}|\rho_{0})$, of
$\mu_{\rho_{0}}$-measurable subsets of $\mathcal{X}$ we can
construct the family $\{\mathcal{X}'_{i}\}$ consisting of the same
number of Borel subsets of $\mathcal{X}$ such that
$\mu_{\rho_{0}}((\mathcal{X}'_{i}\setminus\mathcal{X}_{i})\cup(\mathcal{X}_{i}\setminus\mathcal{X}'_{i}))=0\,$
and $\,\bigcup_{i}\mathcal{X}'_{i}=\mathcal{X}$. For each $i$ the
state
$$
\sigma_{0}^{i}=\frac{1}{\nu_{0}(\mathcal{B}_{i}^{m_{0}})}
\int_{\mathcal{B}_{i}^{m_{0}}}\rho
\nu_{0}(d\rho)=\frac{1}{\mu_{\rho_{0}}(\mathcal{X}'_{i})}
\int_{\mathcal{X}'_{i}}\sigma(x|\rho_{0})
\mu_{\rho_{0}}(dx)=\frac{\mathfrak{M}(\mathcal{X}'_{i})
[\rho_{0}]}{\operatorname{Tr}\mathfrak{M}(\mathcal{X}'_{i})[\rho_{0}]}
$$
is the posteriori state, corresponding to the set
of outcomes $\mathcal{X}'_{i}$ and the a priori state $\rho_{0}$.

For each $i$ let
$\sigma_{n}^{i}=\frac{\mathfrak{M}(\mathcal{X}'_{i})[\rho_{n}]}
{\operatorname{Tr}\mathfrak{M}(\mathcal{X}'_{i})[\rho_{n}]}$ be the
posteriori state, corresponding to the set of
outcomes $\mathcal{X}'_{i}$ and the a priori state $\rho_{n}$.\footnote{Since
$\operatorname{Tr}\mathfrak{M}(\mathcal{X}'_{i})[\rho_{0}]=\mu_{\rho_{0}}(\mathcal{X}'_{i})>0$
the state $\sigma_{n}^{i}$ is correctly defined for all sufficiently
large $n$.} By lower semicontinuity of the function $f$ and since
$\lim_{n\rightarrow+\infty}\mathfrak{M}(\mathcal{X}'_{i})[\rho_{n}]=\mathfrak{M}(\mathcal{X}'_{i})[\rho_{0}]$
we have
\begin{equation}
\label{eq31}
\sum_{i}\mu_{\rho_{n}}(\mathcal{X}'_{i})f(\sigma_{n}^{i})\ge
\sum_{i}\mu_{\rho_{0}}(\mathcal{X}'_{i})f(\sigma_{0}^{i})-\frac{1}{3}\Delta
\end{equation}
for all sufficiently large $n$.

By Jensen's inequality (Proposition~\ref{pA-2}) convexity and lower
semicontinuity of the function $f$ implies
\begin{equation}
\label{eq32}
\mu_{\rho_{n}}(\mathcal{X}'_{i})f(\sigma_{n}^{i})\le
\int_{\mathcal{X}'_{i}}f(\sigma(x|\rho_{n}))\mu_{\rho_{n}}(dx)
\qquad
\forall\, i,n.
\end{equation}
By using \eqref{eq30}--\eqref{eq32} we obtain
\begin{align*}
\int_{\mathcal{X}}f(\sigma(x|\rho_{n}))\mu_{\rho_{n}}(dx)
&=
\sum_{i}\int_{\mathcal{X}'_{i}}f(\sigma(x|\rho_{n}))\mu_{\rho_{n}}(dx)\ge
\sum_{i}\mu_{\rho_{n}}(\mathcal{X}'_{i})f(\sigma_{n}^{i})
\\
&\ge
\sum_{i}\mu_{\rho_{0}}(\mathcal{X}'_{i})f(\sigma_{0}^{i})-\frac{1}{3}\Delta
\ge
\int_{\mathfrak{S}(\mathcal{H})}f(\rho)\nu_{0}(d\rho)-\frac{2}{3}\Delta
\end{align*}
for all sufficiently large $n$, which contradicts to \eqref{eq29}.
\end{proof}

The author is grateful to A.S.Holevo for the help and useful
discussion. The author is also grateful to the referees for useful
remarks.

\end{fulltext}

\end{document}